\documentclass[11pt, draftcls, letterpaper, onecolumn]{IEEEtran}
\usepackage{cite}
\ifCLASSINFOpdf
   \usepackage[pdftex]{graphicx}
   \DeclareGraphicsExtensions{.pdf,.jpeg,.png}
\else
   \usepackage[dvips]{graphicx}
   \DeclareGraphicsExtensions{.eps}
\fi
\usepackage[cmex10]{amsmath}
\begin{document}
%\title{Scenario-Adaptive Sphere Decoding for Complexity Efficient MIMO Iterative Receivers}
\title{Approximate MIMO Iterative Processing with Adjustable Complexity Requirements}
\author{Konstantinos~Nikitopoulos,~\IEEEmembership{Member,~IEEE,} and~Gerd~Ascheid,

~\IEEEmembership{Senior Member,~IEEE}

%\author{Konstantinos~Nikitopoulos,~\IEEEmembership{Member,~IEEE,}~Dan~Zhang,~Gerd~Ascheid,
%~\IEEEmembership{Senior Member,~IEEE,}

%and~Heinrich~Meyr,~\IEEEmembership{Life Fellow,~IEEE}% <-this % stops a space
\thanks{The authors are with the Faculty of Electrical Engineering and Information Technology, RWTH Aachen University, Aachen, Germany (e-mail: Konstantinos.Nikitopoulos@iss.rwth-aachen.de).}
\thanks{This work has been supported by the UMIC Research Center, RWTH Aachen University.}}% <-this % stops a space
%\thanks{Manuscript received April 19, 2005; revised January 11, 2007.}}

%\markboth{To be submitted to IEEE Transactions on Wireless Communications}%
%{Nikitopoulos\MakeLowercase{\textit{et al.}}: Scenario-Adaptive Soft-Input, Soft-Output Sphere Decoding}
\markboth{The final version of this paper appears in IEEE Transactions on Vehicular Technology}{}




\maketitle
\vspace{-2cm}
\begin{abstract}
%\boldmath
Targeting always the best achievable bit error rate (BER) performance in iterative receivers operating over multiple-input multiple-output (MIMO) channels may result in significant waste of resources, especially when the achievable BER is orders of magnitude better than the target performance (e.g., under good channel conditions and at high signal-to-noise ratio (SNR)). In contrast to the typical iterative schemes, a practical iterative decoding framework that approximates the soft-information exchange is proposed which allows reduced complexity sphere and channel decoding, adjustable to the transmission conditions and the required bit error rate. With the proposed approximate soft information exchange the performance of the exact soft information can still be reached with significant complexity gains.

\end{abstract}

\begin{IEEEkeywords}
MIMO systems, iterative methods, soft-input soft-output detection, sphere decoding
%IEEEtran, journal, \LaTeX, paper, template.
\end{IEEEkeywords}

% \fi
\IEEEpeerreviewmaketitle

\section{Introduction}
\IEEEPARstart{M}{ultiple}-input, multiple-output (MIMO) transmission with iterative receiver
processing of soft information has been proposed as a very efficient way to achieve
near-capacity transmission at the cost of a highly increased computational
effort \cite{Hochwald03}, \cite{Vikalo04}. These increased processing requirements prevent practical implementations from meeting the
theoretical performance limits due to the resulting increased energy consumption and the increased latency requirements.

Although the best achievable performance is required for transmission over ``unfavorable" transmission environments (i.e., ill-conditioned transmission channels and in the low SNR regime) and for increased performance (in terms of BER) it may be unnecessary for transmission over a good channel and/or reduced performance needs. For the \emph{non}-iterative case and in order to avoid the highly complex optimal solution when it is not required, receivers supporting both optimal and suboptimal algorithmic solutions (e.g., zero-forcing, MMSE) have been proposed \cite{Lai10}.
%Then, the receiver chooses the less complex of those which can provide the required performance.
However, such approaches impose an increased (silicon) area occupation
%which increases with the number of supported algorithms.
and involve tedious selection processes in order to choose the appropriate algorithm from the set of the available ones. These selection processes typically demand \emph{performance prediction} methods. Therefore, the applicability of such methods is restricted to those scenarios where performance prediction is available and where the selection strategies are computationally efficient (i.e., they can be performed with low processing overhead). Consequently, such approaches are not convenient for iterative systems where performance prediction (per iteration) is very difficult to be acquired, especially when sub-optimal algorithms are involved in the iterative process.

The proposed scheme targets the avoidance of the unnecessary processing which would further increase the reliability of those bits which reach the required performance (i.e., meet the TER) at early iterations of the decoding process. This simplification is performed only when the \emph{convergence} of the iterative process is not expected to be significantly affected (both in terms of convergence point and convergence rate).
Instead of supporting several soft demapping algorithms the proposed iterative decoding framework employs a single, flexible soft demapper which is efficiently realized in terms of sphere decoding (SD) and it can adjust its complexity to the given channel scenario and to the required target bit error rate (TER) performance. Additional savings can be achieved at the channel decoder side since the proposed framework allows selective decoding.

The proposed approach can be applied to any SD scheme. However, %\cite{Viter93,Damen03,Agrell02,Wang06}
the SD of \cite{SISO_SD} is employed which requires only minor modifications in order to be accommodated at the proposed scheme. Furthermore, this SD can ensure the (exact) \emph{max-log} MAP performance when it is required.
%The soft demapper is one of the main and most computational intensive processing blocks in
%MIMO iterative processing. It is employed to calculate the \emph{a-posteriori} soft information of the received bits and it is typically realized by means of the well-known sphere decoder (SD) \cite{Viter93,Damen03,Agrell02,Wang06} which converts the exhaustive search (aiming at the optimal solution) into a constrained tree search. Two
%main approaches have been proposed in the literature in order to efficiently traverse the corresponding tree and to calculate the required soft information. The first, which is  adopted by this work, can ensure the (exact) \emph{max-log} MAP performance \cite{Wang06},\cite{Studer08}. It performs a depth-first traversal with the Schnorr and Euchner enumeration \cite{Schnorr94} in order to efficiently perform the corresponding tree search in terms of the number of visited nodes. This approach necessitates a tree search operation per iteration in contrast to the \emph{list} sphere decoding approach of \cite{Hochwald03,Vikalo04,Boutros03,Wei09Globecom} where the tree search can take place only at the first iteration at the cost of reduced performance. In addition, large memories are required to save the corresponding lists, the size of which can hardly be optimized to minimize the demapping complexity without compromising the bit error rate (BER) performance in different operational scenarios.
In order to adjust the SD's complexity, the scheme selectively updates only the log-likelihood ratio (LLR) values of the bits whose exact value is required to preserve the TER and the \emph{convergence} properties of the iterative process. Additionally, \emph{performance-driven} LLR clipping is employed to avoid the unnecessary processing which eventually, will result in BER performance better than the TER. The fundamental concept of reducing SD's complexity by bounding the LLR clipping value at the cost of reduced BER performance has been proposed and demonstrated via extensive simulations in \cite{ETH_SS,SISO_SD}. However, the question of how to practically set the LLR clipping value in order to adjust (on-the-fly) the receiver's complexity to the TER requirements has not been addressed. In this context, several approaches of different efficiency are discussed for relating the LLR clipping value to the TER. The proposed approaches do not require the exact relationship between the LLR clipping value and the resulting performance. Therefore, they are generally applicable to any kind of transmission scenario (i.e., channel, SNR, coding scheme, etc.) without demanding extensive simulations and/or any tedious and computationally intensive \emph{performance-prediction} methods which should account for all the performance-affecting parameters and all the possible transmission scenarios.

The paper is organized as follows. In Section II the typical soft-input, soft-output (SISO) processing for transmission over MIMO channels is outlined and the basic observations which are explored by the proposed approach are made. In Section III the details of the modified, approximate, iterative processing are presented together with the required modifications to the SD and the channel decoder. Complexity issues are also discussed. Finally, in Section IV, the proposed approach is validated and the corresponding complexity gains are depicted via extensive simulations.

%However, it is shown that prior knowledge on the achievable system performance (i.e., performance %prediction) per per iteration could further reduce the SD's complexity.



%\hfill January 11, 2007
\section{Soft-Input, Soft-Output Receiver Processing for MIMO Systems}
Typically, as shown in Fig. 1, during the $q$-th iteration and over several MIMO channel utilizations, the soft-demapper module employs the corresponding \emph{a-priori} $\mathbf{L}_{A}^{(q)}$ soft information vector (nulled for $q=0$) and the related received vectors ${\bf{y}}$ in order to calculate the \emph{a-posteriori} or \emph{intrinsic} $\mathbf{L}_{D}^{(q)}$ as well as the \emph{extrinsic} $\mathbf{L}_{E}^{(q)}=\mathbf{L}_{D}^{(q)}-\mathbf{L}_{A}^{(q)}$ soft information of the coded bits. The \emph{extrinsic} soft information is then de-interleaved and fed to the SISO channel decoder as \emph{a-priori} information $\mathbf{\tilde{L}}_{A}^{(q)}$ in order to calculate channel decoder's \emph{a-posteriori} soft information $\mathbf{\tilde{L}}_{D}^{(q)}$. This can be used by an early-stopping module to evaluate the resulted error rate performance, to be discussed later in detail. Then, if the TER performance has been reached, the iterations stop and hard decisions will be made based on $\mathbf{\tilde{L}}_{D}^{(q)}$. Otherwise, the decoder's \emph{extrinsic} information is calculated  ($\mathbf{\tilde{L}}_{E}^{(q)}=\mathbf{\tilde{L}}_{D}^{(q)}-\mathbf{\tilde{L}}_{A}^{(q)}$) and after being interleaved it is fed to the soft-demapper module as \emph{a-priori} $\mathbf{L}_{A}^{(q+1)}$ information to be used during the next iteration.

\subsection{Soft Demapping in Terms of Sphere Decoding}
In MIMO transmission with  ${M_T}$ transmit and ${{M}_{R}}\ge {{M}_{T}}$ receive antennas, at the $u$-th MIMO channel utilization, the interleaved coded bits are grouped into blocks ${B_{t,u}}$ ($t = 1,...,{M_T}$ and $u = 1,...,{U}$ with ${U}$ being the number of channel utilizations per code block) in order to be mapped onto symbols  ${s_{t,u}}$ of a constellation set $S$ of cardinality $\left| S \right|$. The bipolar $k$-th bit resides in block ${B_{{\left\lceil {{k \mathord{\left/
 {\vphantom {k {{{\log }_2}\left| S \right|}}} \right.\kern-\nulldelimiterspace} {{{\log }_2}\left| S \right|}}} \right\rceil },u}}$ and the blocks ${B_{t,u}}$ are mapped onto the symbols ${s_{t,u}}$  by a given mapping function (e.g., Gray mapping). The corresponding received ${M_R}\times{1}$ vector ${\bf{y}}_{u}$ is, then, given by
\begin{equation}
{\bf{y}}_{u} = {\bf{H}}_{u}{\bf{s}}_{u} + {\bf{n}}_{u},	
\end{equation}
where ${\bf{H}}_{u}$ is the ${M_R}\times{M_T}$ complex channel matrix and ${\bf{s}}_{u} = {\left[ {{s_{1,u}},{s_{2,u}},...,{s_{{M_T,u}}}} \right]^T}$ is the transmitted symbol vector. Then, $c_{b,i,u}$ is the $b$-th bit of the $i$-th entry of ${\bf{s}}_{u}$ and the term ${\bf{n}}_{u}$
 is the noise vector, consisting of i.i.d., zero-mean, complex, Gaussian  samples with variance $2\sigma _n^2$.

 As already noted, the role of the soft demapper is to calculate at every iteration the \emph{a-posteriori} \emph{log-likelihood} ratios (LLRs) for all the symbols residing in the frame to be decoded. Namely, it calculates
 \begin{equation}
{{L}_{D}}\left( {{c}_{b,i,u}} \right)=\ln \left( \frac{P[{{c}_{b,i,u}}=+1|{{\mathbf{y}}_{u}},{{\mathbf{H}}_{u}}]}{P[{{c}_{b,i,u}}=-1|{{\mathbf{y}}_{u}},{{\mathbf{H}}_{u}}]} \right),~\forall b,i,u.
 \end{equation}
Assuming that the corresponding bits are statistically independent (due to interleaving) and by employing the Bayes' theorem (2) can be expressed as
 % \begin{equation}
% \[{L_{D}}\left( {{c_{b,i,u}}} \right) = \ln \left( {\sum\limits_{{\bf{s}}_{u} \in S_{b,i,u}^{ + 1}} {p({{\bf{y}}_u}|{{\bf{s}}_u},{{\bf{H}}_u})P[{{\bf{s}}_u}]} } \right)\]
 \begin{equation}
 {L_{D}}\left( {{c_{b,i,u}}} \right) = \ln \left( {\sum\limits_{{\bf{s}}_{u} \in S_{b,i,u}^{ + 1}} {p({{\bf{y}}_u}|{{\bf{s}}_u},{{\bf{H}}_u})P[{{\bf{s}}_u}]} } \right)
 - \ln \left( {\sum\limits_{{\bf{s}}_{u} \in S_{b,i,u}^{ - 1}} {p({{\bf{y}}_u}|{{\bf{s}}_u},{{\bf{H}}_u})P[{{\bf{s}}_u}]} } \right)
  \end{equation}
 where
   \begin{equation}
   p({{\bf{y}}_u}|{{\bf{s}}_u},{{\bf{H}}_u}) = \frac{1}{{{{\left( {2\pi \sigma _n^2} \right)}^{{M_R}}}}}\exp \left( { - \frac{{{{\left\| {{{\bf{y}}_u} - {{\bf{H}}_u}{{\bf{s}}_u}} \right\|}^2}}}{{2\sigma _n^2}}} \right)
   \end{equation}
 and $S_{b,i,u}^{ \pm 1}$ are the sub-sets of possible ${{\bf{s}}_u}$ symbol sequences having the $b$-th bit value of their $i$-th ${\bf{s}}_{u}$ entry equal to $ \pm 1$, while $P[{{\bf{s}}_u}]$ is the available \emph{a-priori} information provided by the channel decoder at the previous iteration. However, in order to compute (3) exhaustive calculations over all possible ${\bf{s}}_{u}$ symbols are required which leads to prohibitive computational complexity especially for large $M_T$ values. This problem is typically tackled by employing the standard \emph{max-log} approximation and by QR decomposition of the MIMO channel matrix \cite{Hochwald03}. Then the problem transforms into an equivalent tree-search which can be efficiently solved by means of sphere decoding. In detail, the channel matrix ${\bf{H}}_{u}$ can be QR decomposed into ${\bf{H}}_{u} = {\bf{Q}}_{u}{\bf{R}}_{u}$, with ${\bf{Q}}_{u}$ a unitary \mbox{${M_R}\times{M_T}$} matrix and ${\bf{R}}_{u}$ an ${M_T}\times{M_T}$ upper triangular matrix with elements ${R_{i,j,u}}$ and real-valued positive diagonal entries.
 Then, under the \emph{max-log} approximation the LLR calculation problem can be trasformed to \cite{Hochwald03,SISO_SD}
\begin{equation}
{{L}_{D}}\left( {{c}_{b,i,u}} \right) \approx \mathop {\min }\limits_{{\bf{s}}_{u} \in S_{b,i,u}^{ - 1}} \left\{ {I({\bf{s}}_{u})} \right\} - \mathop {\min }\limits_{{\bf{s}}_{u} \in S_{b,i,u}^{ + 1}} \left\{ {I({\bf{s}}_{u})} \right\}\
\end{equation}
where $I({\bf{s}}_{u}) = {I_{channel}}({\bf{s}}_{u}) + {I_{prior}}({\bf{s}}_{u})$, with

%\[{I_{channel}}({\bf{s}}_{u}) = \frac{1}{2\sigma _{n}^{2}}{{\left\| \mathbf{y}{{'}_{u}}-{{\mathbf{R}}_{u}}{{\mathbf{s}}_{u}} \right\|}^{2}}=\]
\begin{equation}
{I_{channel}}({\bf{s}}_{u}) = \frac{1}{2\sigma _{n}^{2}}{{\left\| \mathbf{y}{{'}_{u}}-{{\mathbf{R}}_{u}}{{\mathbf{s}}_{u}} \right\|}^{2}}=
\frac{1}{{2\sigma _n^2}}\sum\limits_{l = 1}^{{M_T}} {{{\left| {y{'_{l,u}} - \sum\limits_{j = i}^{{M_T}} {{R_{l,j,u}}{s_{j,u}}} } \right|}^2}}
\end{equation}
being the \emph{channel-based} part of the soft information, with  ${\bf{y'}}_{u} = {{{\bf{Q}}_{u}}^H}{\bf{y}}_{u} = {\left[ {y{'_{1,u}},y{'_{2,u}},...,y{'_{{{M_T,u}}}}} \right]^T}$ while, for statistically independent symbols
\begin{equation}
{I_{prior}}({\bf{s}}_{u}) = - \ln P({\bf{s}}_{u})= - \sum\limits_{l = 1}^{{M_T}} {\ln P\left[ {{s_{l,u}}} \right]}
\end{equation}
 being the \emph{a-priori} part of the soft information, which is always non-negative. Eq. (5) shows how the \emph{max-log} LLR calculation can be reformulated into two constrained minimization problems per decoded bit over the different symbol-vector subsets (i.e., $S_{b,i,u}^{ \pm 1}$).

Since $P[c]$ is related to its corresponding LLR as
\begin{equation}
P\left[ c \right]=\frac{\exp \left( -{\left| {{L}}\left( c \right) \right|}/{2}\; \right)}{1+\exp \left( -\left| {{L}}\left( c \right) \right| \right)}\exp \left( {c{{L}}\left( c \right)}/{2}\; \right)
\end{equation}
and after extracting the mutually exclusive terms from the two minimization problems of (5), (7) becomes
\begin{equation}
{{I}_{prior}}({{\mathbf{s}}_{u}})=\sum\limits_{l=1}^{{{M}_{T}}}{\sum\limits_{j=1}^{{{\log }_{2}}\left| S \right|}{\frac{1}{2}}}\left( \left| {{L}_{A}}\left( {{c}_{j,l,u}} \right) \right|-{{c}_{j,l,u}}{{L}_{A}}\left( {{c}_{j,l,u}} \right) \right)
\end{equation}
without affecting the optimality of (5). The \emph{extrinsic} information is then calculated by subtracting the \emph{a-priori} from the \emph{a-posteriori} information and, after de-interleaving, it is fed to the SISO channel decoder as \emph{a-priori} information
\[
{{\tilde{L}}_{A}}\left( {{{\tilde{c}}}_{k}}={{\pi }^{-1}}\left( {{c}_{b,i,u}} \right) \right)={{L}_{E}}\left( {{c}_{b,i,u}} \right)
={{L}_{D}}\left( {{c}_{b,i,u}} \right)-{{L}_{A}}\left( {{c}_{b,i,u}} \right)
\]
%\[
%={{L}_{D}}\left( {{c}_{b,i,u}} \right)-{{L}_{A}}\left( {{c}_{b,i,u}} \right)
%\]
with $k=1,...,K$ and $K={U}{M_{T}}{{\log }_{2}}\left| S \right|$. Then,
\begin{equation}
{{L}_{E}}\left( {{c}_{b,i,u}} \right)\approx \mathop {\min }\limits_{{\bf{s}}_{u} \in S_{b,i,u}^{ - 1}} \left\{ {I'({\bf{s}}_{u})} \right\} - \mathop {\min }\limits_{{\bf{s}}_{u} \in S_{b,i,u}^{ + 1}} \left\{ {I'({\bf{s}}_{u})} \right\}\
\end{equation}
with $I'({\bf{s}}_{u}) = {{I}_{channel}}({\bf{s}}_{u}) + {{I'}_{prior}}({\bf{s}}_{u})$, and
\begin{equation}
I{{'}_{prior}}({{\mathbf{s}}_{u}})=\sum\limits_{l=1}^{{{M}_{T}}}{\sum\limits_{j=1(j\ne b;l=i)}^{{{\log }_{2}}\left| S \right|}{\left( \left| {{L}_{A}}\left( {{c}_{j,l,u}} \right) \right|-{{c}_{j,l,u}}{{L}_{A}}\left( {{c}_{j,l,u}} \right) \right)}}
\end{equation}

 By inspecting the equations above, some basic observations can be made which will later be exploited by the proposed approach. As it can be seen from (6) and (9), even if ${I}_{channel}$  significantly affects the solution of the minimization problem of (5), it does not vary over iterations in unlike to ${I}_{prior}$. In addition, the strongly unlikely bits (i.e., the ones of opposite sign to ${L}_{A}$ with high $\left| {{L}_{A}} \right|$ which result in low $P[c]$, see (8)) contribute with high ${I}_{prior}$ values (see (9)). Thus, the symbol-vectors consisting of such bits can be assumed to be weak candidate solutions for the minimization problems of (5). On the other hand, the highly likely solutions (of the same sign with ${L}_{A}$) contribute with zero ${I}_{prior}$. Therefore, the symbol-vector solutions are expected to consist of the symbols having the smallest ${I}_{channel}$ among the ones with small ${I}_{prior}$ values. Namely, among the symbols consisting of highly likely (i.e., of the same sign with ${L}_{A}$) and loosely unlikely bits (i.e., of low $\left| {{L}_{A}} \right|$ and opposite sign than ${L}_{A}$). Since for high $\left| {{L}_{A}} \right|$ value the bit with the opposite sign to ${{L}_{A}}$ is not expected to belong to the symbol-vector solutions, while the bit with the same sign to ${{L}_{A}}$ contributes with zero ${I}_{prior}$ independently of the $|{{L}_{A}}|$ value, it can be concluded that approximate calculation of the related strong (i.e., of high amplitude) \emph{a-priori} information is not expected to significantly affect SD performance.

  An additional property related to the soft-information flow is that since ${I}_{channel}$ remains constant over iterations the value of  ${{L}_{E}}\left( {{c}_{b,i,u}} \right)$ will also be constant as long as the symbol-vector solutions of (5) do not vary over iterations and consist of  (highly or loosely) likely bits contributing with zero ${I}_{prior}$, even if the \emph{a-priori} information varies. This attribute is expected over later iterations when (and if) the corresponding symbol solutions are dominated by the highly likely symbols. Then, instead of recalculating their corresponding ${{L}_{D}}$ and ${{L}_{E}}$ values the ones of the previous iteration can be employed. This property can be exploited by any SD approach to provide computational complexity gains. These gains are expected to increase with the number of iterations where the average reliability of the bits is expected to increase \cite{tenBrink01}.

\subsection{SISO Channel Decoding}
Similarly to the soft demapper, after de-interleaving, the corresponding input soft-information is exploited in order to calculate the corresponding \emph{a-posteriori} information, as
\begin{equation}
{{\tilde{L}}_{D}}\left( {{{\tilde{c}}}_{k}} \right)=\ln \left( \frac{P[{{\tilde{c}}_{k}}=+1|{{{\mathbf{\tilde{L}}}}_{A}}}{P[{{\tilde{c}}_{k}}=-1|{{{\mathbf{\tilde{L}}}}_{A}}} \right)
\end{equation}
where for $\mathbf{\tilde{c}}$ being the encoded sequence after de-interleaving, it is expressed as
\[{{\tilde{L}}_{D}}\left( {{{\tilde{c}}}_{k}} \right)=\ln \left( \sum\limits_{\mathbf{\tilde{c}}:\tilde{C}_{k}^{+1}}{P\left[ \mathbf{\tilde{c}}|{{{\mathbf{\tilde{L}}}}_{A}} \right]} \right)-\ln \left( \sum\limits_{\mathbf{\tilde{c}}:\tilde{C}_{k}^{-1}}{P\left[ \mathbf{\tilde{c}}|{{{\mathbf{\tilde{L}}}}_{A}} \right]} \right)=\]
%\[=\ln \left( \sum\limits_{\mathbf{\tilde{c}}:\tilde{C}_{k}^{+1}}{\exp \sum\limits_{i=1}^{K}{\ln P\left[ {{\tilde{c}}_{i}}|{{{\tilde{L}}}_{A}}\left( {{\tilde{c}}_{i}} \right) \right]}} \right)-\]
\begin{equation}
=\ln \left( \sum\limits_{\mathbf{\tilde{c}}:\tilde{C}_{k}^{+1}}{\exp \sum\limits_{i=1}^{K}{\ln P\left[ {{\tilde{c}}_{i}}|{{{\tilde{L}}}_{A}}\left( {{\tilde{c}}_{i}} \right) \right]}} \right)-
\ln \left( \sum\limits_{\mathbf{\tilde{c}}:\tilde{C}_{k}^{-1}}{\exp \sum\limits_{i=1}^{K}{\ln P\left[ {{\tilde{c}}_{i}}|{{{\tilde{L}}}_{A}}\left( {{\tilde{c}}_{i}} \right) \right]}} \right)
\end{equation}
with $\tilde{C}_{k}^{\pm 1}$ being the set of bit sequences $\mathbf{\tilde{c}}$ with their $k$-th bit equal to $\pm 1$. Then, (13) can be efficiently calculated by the well-known BCJR-MAP algorithm \cite{BCJR}, \cite{BenedettoII}.

Similar observations with those holding for the soft-demapper can be made for the SISO outer channel decoder. In detail, (13) shows that the most significant contributing sequences $\mathbf{\tilde{c}}$ to the ${{\tilde{L}}_{D}}$ calculation are those with their non-positive $\sum\limits_{k=1}^{K}{\ln P\left[ {{{\tilde{c}}}_{i}}|{{{\tilde{L}}}_{A}}\left( {{{\tilde{c}}}_{i}} \right) \right]}$ terms being close to zero, or equivalently, sequences which do not contain highly unlikely bits (of very low $P\left[ {{{\tilde{c}}}_{i}}|{{{\tilde{L}}}_{A}}\left( {{{\tilde{c}}}_{i}} \right) \right]$). Additionally, under the approximation of \cite{Baero03}
\begin{equation}
\ln P\left( {{{\tilde{c}}}_{k}}|{{{\tilde{L}}}_{A}}\left( {{{\tilde{c}}}_{k}} \right) \right)\approx \frac{1}{2}\left( {{{\tilde{c}}}_{k}}{{{\tilde{L}}}_{A}}\left( {{{\tilde{c}}}_{k}} \right)-\left| {{{\tilde{L}}}_{A}}\left( {{{\tilde{c}}}_{k}} \right) \right| \right)
\end{equation}
which holds for large $|{{\tilde{L}}_{A}}\left( {{\tilde{c}}_{k}} \right)|$ values (typically larger than 2), it can be deduced that for (highly) likely bits the terms $\ln P\left[ {{{\tilde{c}}}_{i}}|{{{\tilde{L}}}_{A}}\left( {{{\tilde{c}}}_{i}} \right) \right]$ equal zero independently of the exact ${{{\tilde{L}}}_{A}}$ value. Therefore, similar to the SD, approximate  calculation of the strong soft information, (i.e., of high $|{{\tilde{L}}_{A}}\left( {{\tilde{c}}_{k}} \right)|$) is not expected to significantly affect the outcome of the SISO channel decoder.

By using (13) we can express the \emph{extrinsic} information of the outer SISO decoder as
\[{{\tilde{L}}_{E}}\left( {{{\tilde{c}}}_{k}} \right)={{\tilde{L}}_{D}}\left( {{{\tilde{c}}}_{k}} \right)-{{\tilde{L}}_{A}}\left( {{{\tilde{c}}}_{k}} \right)=\]
%\[=\ln \left( \sum\limits_{\mathbf{c}:C_{k}^{+1}}{\exp \sum\limits_{i=1,i\ne k}^{K}{\ln P\left[ {{c}_{i}}|{{{\tilde{L}}}_{A}}\left( {{c}_{i}} \right) \right]}} \right)-\]
\begin{equation}
=\ln \left( \sum\limits_{\mathbf{c}:C_{k}^{+1}}{\exp \sum\limits_{i=1,i\ne k}^{K}{\ln P\left[ {{c}_{i}}|{{{\tilde{L}}}_{A}}\left( {{c}_{i}} \right) \right]}} \right)-
\ln \left( \sum\limits_{\mathbf{c}:C_{k}^{-1}}{\exp \sum\limits_{i=1,i\ne k}^{K}{\ln P\left[ {{c}_{i}}|{{{\tilde{L}}}_{A}}\left( {{c}_{i}} \right) \right]}} \right)
\end{equation}
from which it becomes apparent that the \emph{extrinsic} information is a function of the soft information of all other bits different than ${{\tilde{c}}_{k}}$.
Practically, those bits which mainly affect the calculation of ${{\tilde{L}}_{D}}\left( {{\tilde{c}}_{k}} \right)$, here denoted as ${{\Lambda }_{{{\tilde{c}}_{k}}}}$, are the ones residing in a region around ${\tilde{c}}_{k}$ of a size related to the constraint length of the code \cite{Wu09}.
 Thus, a high $|{{\tilde{L}}_{E}}\left( {{{\tilde{c}}}_{k}} \right)|$ value denotes that the \emph{a-priori} information of the surrounding ${{\Lambda }_{{{\tilde{c}}_{k}}}}$ bits can introduce such high reliability to the bit that it finally becomes strongly reliable even if its own \emph{a-priori} information is loose (since ${{\tilde{L}}_{D}}\left( {{{\tilde{c}}}_{k}} \right)={{\tilde{L}}_{E}}\left( {{{\tilde{c}}}_{k}} \right)+{{\tilde{L}}_{A}}\left( {{{\tilde{c}}}_{k}} \right)$). Therefore, it will later be assumed that $|{{\tilde{L}}_{E}}\left( {{{\tilde{c}}}_{k}} \right)|$
is a good indicator of the \emph{correcting capabilities} of the outer code on the specific bit.

Since ${{\tilde{L}}_{E}}\left( {{{\tilde{c}}}_{k}} \right)$  is a function of the ``new" soft-information content calculated at the soft-demapper side (${{L}_{A}}\left( {{\Lambda }_{{{\tilde{c}}_{k}}}} \right)={{L}_{E}}\left( \pi ({{\Lambda }_{{{\tilde{c}}_{k}}}}) \right)$), it is also assumed to be a good indicator of the \emph{convergence rate} of the corresponding bit. Therefore, in the sequel, when the \emph{extrinsic} information of a bit is loose, even if its \emph{a-priori} information is strong (which will lead to strong \emph{a-posteriori} information), its soft information exchange will not be approximated in order to preserve its convergence rate. In other words, loose \emph{extrinsic} information for ${{\tilde{c}}_{k}}$ indicates loose information content for (at least some of) the bits belonging to ${{\Lambda }_{{{\tilde{c}}_{k}}}}$. Therefore, since the decoding quality of the ${{\Lambda }_{{{\tilde{c}}_{k}}}}$ bits is also affected by the soft information content of ${{\tilde{c}}_{k}}$ (since ${{\tilde{c}}_{k}}$  belongs to the theirs ${\Lambda }$ region: ${{\tilde{c}}_{k}}\in {{\Lambda }_{\left( {{\Lambda }_{{{{\tilde{c}}}_{k}}}} \right)}}$) approximating its soft information may affect the convergence of the ${{\Lambda }_{{{\tilde{c}}_{k}}}}$ bits. At this point the following, additional relation can be brought up. The overall convergence rate of the iterative process is dominated by the slowest converging bit. Thus, approximating the soft information of the fast converging bits (or equivalently the ones with the strong \emph{a-priori} information to be fed to the following module) does not significantly affect convergence. In contrast, the convergence properties can be affected by approximating the soft information of the slowest converging bits (as we already discussed in Sections II.A and B).

Finally, it is noted that when ${{\Lambda }_{{{\tilde{c}}_{k}}}}$ have reached the state of constant information exchange (i.e., over later iterations) ${{\tilde{L}}_{E}}\left( {{{\tilde{c}}}_{k}} \right)$ will be also constant.

\subsection{Performance Driven Early Iteration Stopping}
In practical iterative schemes frame-based, early-stopping mechanisms are typically employed to reduce the average required number of iterations without compromising the resulting performance. In such mechanisms, the convergence status is checked by means of a specific pre-selected criterion for terminating iterations. Two main classes of criteria have been proposed for this purpose. The first class is based on cross-entropy metrics \cite{Hagenauer96,Shao99,Costello04} and it is typically employed to identify if a frame still converges over iterations. The second class \cite{Hagenauer96,Land01,Letzepis03,Strinati07} employs, directly, the calculated \emph{a-posteriori} LLR values to \emph{evaluate} the already achieved error rate performance, and then terminate iterations when the TER is achieved. Even if the cross-entropy methods have been shown to be more efficient in identifying if an iterative system is still converging, in this work the second approach is employed since it links the number of iterations directly to the error-rate performance. Thus, it can  early terminate the iteration process when the TER is reached, even if the iterative system still converges.

According to \cite{Land00} the BER of the decoded block can be evaluated after SISO decoding as
\begin{equation}
{{\hat{P}}_{b}}=\frac{1}{{{N}_{I}}}\sum\limits_{i=1}^{{{N}_{I}}}{\frac{1}{1+\exp \left( \left| {{{\tilde{L}}}_{D}}\left( \tilde{c}_{i}^{I} \right) \right| \right)}}
\end{equation}
where $\tilde{c}_{i}^{I}$ are the $N_{I}$ information bits. From the above equation some simple conclusions can be drawn which will be exploited by the proposed approach. The provided BER, as well as the corresponding estimate, are expected to be dominated by the bits with small $\left| {{{\tilde{L}}}_{D}}\left( \tilde{c}_{i}^{I} \right) \right|$ values. Thus, for reliable error rate prediction and in order to preserve the convergence behavior of the iterative process, no approximation is attempted for the weak $\left| {{{\tilde{L}}}_{D}}\left( \tilde{c}_{i}^{I} \right) \right|$ values. In the same context, if all contributing terms in (16) with BER$\le$TER (or equivalently with $\left| {{{\tilde{L}}}_{D}}\left( \tilde{c}_{i} \right) \right|\le {{\tilde{L}}_{TER}}=\ln (TE{{R}^{-1}}-1)$) are accurately calculated while the others are clipped to a value not smaller than ${{\tilde{L}}_{TER}}$ no significant error rate performance degradation is expected for the SNR regimes of achievable performance lower than the TER. However, when clipping is applied, special consideration has to be taken so that the system's convergence is not affected. In this framework, several clipping approaches have been considered (see Section III.B).
It is significant to note that the proposed approximate information flow does not demand any early-stopping mechanism and it does not depend on the choice of the adopted criterion. However, it is considered in this work in order to make meaningful performance comparisons, since such mechanisms are anticipated in practical iterative systems able of adjusting their complexity to the transmission scenario (e.g., SNR) and the TER.

\section{Iterative Receiver Processing of Approximate Soft Information}
\subsection{Approximate Soft Information Flow}
The proposed approximate soft information flow is depicted in Fig. 1. The scheme adjusts its processing requirements to the TER performance by avoiding the unnecessary processing which would further increase the reliability of those bits which are already reliable enough (i.e., meet the TER) at early iterations, but only when such an approximation is not expected to significantly affect the \emph{convergence} properties of the iterative process. Since the \emph{extrinsic} information has been discussed to be a good indicator of both the per-bit correcting capabilities of the channel code and of the per-bit convergence rate (Section II.B) it is employed to decide when the soft information of a specific bit can be safely approximated. In detail, the proposed approach consists of the following steps:
\begin{enumerate}
\item
\emph{Identification of Reliable and Well Converging (RWC) Bits.}

 Using the SISO decoder output at the $q$-th iteration and after the stopping-control check, the reliable bits (i.e., those who meet the TER requirement) with high convergence rate are identified since, more likely, their exact soft-information calculation is not required. In order to characterize the bits their \emph{a-posteriori} and \emph{extrinsic} information is used as described in Section II.B. Consequently, a flag sequence is introduced to indicate the bits whose \emph{a-posteriori} and \emph{extrinsic} information (which is also the \emph{a-priori} information for the soft demapper) is larger than ${{\tilde{L}}_{TER}}$, with
 ${G}^{(q)}\left( {{k}} \right)=1$ when both $\left| {{{\tilde{L}}}_{E}}\left( {{{\tilde{c}}}_{k}} \right) \right|$ and $\left| {{{\tilde{L}}}_{D}}\left( {{{\tilde{c}}}_{k}} \right) \right|$ are larger than ${{\tilde{L}}_{TER}}$ and ${G}^{(q)}\left( {{k}} \right)=0$ otherwise.

When a bit is identified to be a RWC one, it is assumed (perhaps wrongly) that it has reached its constant information flow state. However, if during the next iteration it is not again identified to be an RWC bit, the initial (over the previous iteration) assumption was obviously wrong and it needs to be corrected. In detail, as it has already been discussed in Section II.B, if a bit has been wrongly assumed to be a RWC one (and therefore its soft-information wrongly has not been updated) a negative effect on the convergence characteristics of its neighbor bits is expected (which can also be RWC ones). This negative effect is typically reflected in their \emph{extrinsic} and \emph{a-posteriori} information (see Section II.B). In the same way, affecting the convergence rate of the neighboring bits will affect the soft information of the bit in question (which can become a non-RWC bit).
 Therefore, in order to remedy wrong RWC bit characterizations which could significantly affect the system's convergence, full RWC check over all bits takes place at each iteration.

\item
\emph{Reduced Processing Soft Demapping}

After the RWC bit identification, the flag sequence $\tilde{G}^{(q)}\left( {k} \right)$ is interleaved and the position of the RWC bits at the soft demapper side is identified, then, reduced complexity soft demapping can be performed. For this purpose a slight modification of the SD in \cite{SISO_SD} is described in the next sub-section. There, the SD reduces its processing requirements by skipping the soft information calculation of the RWC bits and by approximately calculating (bounding) the soft-output values of those bits which result in extrinsic information larger than ${{{\tilde{L}}}_{TER}}$. As it will be later discussed in detail, this is only performed when such an approximation is not expected to significantly affect the convergence behavior of the iterative system.

\item
\emph{SISO Decoding with Approximate Soft Information}

Finally, after calculating and de-interleaving the (approximate) \emph{extrinsic} information of the SD, SISO outer decoding follows. For the RWC bits the soft information has not been updated. Therefore, the \emph{a-priori} information of the previous iteration is employed (since constant flow has been assumed).
Then, the processing proceeds with an early-stopping check and an RWC bit update (step 1).

 During the decoding process, no changes are expected on the status of an RWC bit if all its $\Lambda$ bits (i.e., the neighboring bits related to its decoding, see II.B) are also RWC ones. On the contrary, changes may occur whenever in its $\Lambda$ region lie non-RWC bits. Based on this observation, instead of performing full channel decoding, selective decoding can be performed only on the non-RWC bits and their corresponding $\Lambda$ neighbors, resulting in additional complexity gains at the channel decoder side, and will be discussed in detail in Section III.C. In this context, a scenario-adaptive SISO channel decoder may perform selective decoding only on the non-RWC and their surrounding bits belonging to a window of length $w$ centered on each non-RWC bit, so that the non-updated RWC bits do not have any non-RWC ones in their $\Lambda$ region.

 As it will be shown by simulations, the proposed RWC identification is so reliable that no significant changes occur at the state of RCW bits over later iterations, especially for low TER values. Then, the potential gains at the SISO decoder side can be maximized by setting the $w=1$.

\end{enumerate}

 %The proposed SD reduces its complexity requirements by avoiding the processing which will result in $|{{{\tilde{L}}}_{A}}|>{{{\tilde{L}}}_{TER}}$, since for an already reliable and well converging bit the output $|{{{\tilde{L}}}_{D}}|$ will be at least of the same quality (in a worst ${{{\tilde{L}}}_{E}}=0$ case of zero error correcting capability of the code). In reality, the \emph{a-priori} information can be smaller than ${{{\tilde{L}}}_{TER}}$ and still reach TER due to the error correcting capabilities of the code which is manifested as non-zero \emph{extrinsic} information. In this sense, the complexity gains presented in the simulations section could be assumed to be a minimum (i.e., lower floor) of what one could get from such an approach. However, trying to provide tighter RWC bit identification, as a function of the outer code characteristics and the noise and channel statistics, is not trivial and probably results in computational burden which may question the simplicity of the proposed approach.

\subsection{Scenario-Adaptive SD}
 The herein proposed scenario-adaptive SD is based on the typical, depth-first SD approach of \cite{SISO_SD}.
 However, as discussed in Section I, the proposed approximate soft information flow is independent of the soft demapper realization approach.

In detail, in (5) it is shown how the max-log LLR calculation can be reformulated into two constrained  minimization problems over the different symbol-vector subsets (i.e., $S_{b,i,u}^{ \pm 1}$), per decoded bit. For each minimization problem the corresponding tree has its root at level $l = {M_T} + 1$
 and its leafs at level $l = 1$. The ${{I}}\left( {{{\bf{s}}_{u}}} \right)$
 values for any leaf can be calculated recursively by
\begin{equation}
D\left( {{{\bf{s}}_{u}^{(l)}}} \right) =D\left({{{\bf{s}}_{u}^{(l+1)}}} \right) + {{I_{channel}^{(l)}}}\left( {{{\bf{s}}_{u}^{(l)}}} \right) + {{I_{prior}^{(l)}}}\left( {{{\bf{s}}_{u}^{(l)}}} \right)
\end{equation}
where ${{\bf{s}}_{u}^{(l)}} = {[{s_{l,u}},{s_{l+1,u}},...,{s_{{M_T,u}}}]^T}$ are partial symbols vectors, $D\left( {{{\bf{s}}_{u}^{(M_T+1)}}} \right) = 0$,
\begin{equation}
 {{I_{channel}^{(l)}}}\left( {{{\bf{s}}_{u}^{(l)}}} \right) = \frac{1}{{2\sigma _n^2}}{\left| {y{'_{l,u}} - \sum\limits_{j = i}^{{M_T}} {{R_{l,j,u}}{s_{j,u}}} } \right|^2}
\end{equation}
and
\begin{equation}
 {{I_{prior}^{(l)}}}\left( {{{\bf{s}}_{u}^{(l)}}} \right) ={\sum\limits_{j=1}^{{{\log }_{2}}\left| S \right|}{\frac{1}{2}}}\left( \left| {{L}_{A}}\left( {{c}_{j,l,u}} \right) \right|-{{c}_{j,l,u}}{{L}_{A}}\left( {{c}_{j,l,u}} \right) \right)
\end{equation}
with $D\left( {{{\bf{s}}_{u}^{(l)}}} \right)$ being the partial distance (PD) of the ${{{\bf{s}}_{u}^{(l)}}}$ node. Then  ${{I}}\left( {{{\bf{s}}_{u}}} \right)=D\left( {{{\bf{s}}_{u}^{(1)}}} \right)$.

Depth-first tree traversal with Schnorr-Euchner enumeration \cite{Schnorr94} and radius reduction are assumed like in \cite{SISO_SD}. The initial radius is set infinite and whenever a leaf is reached with its corresponding squared radius ${r^2}$ being smaller than $D\left( {{{\bf{s}}_{u}^{(1)}}} \right)$ the ${r^2}$ is updated to $D\left( {{{\bf{s}}_{u}^{(1)}}} \right)$. At each visited node ${{{\bf{s}}_{u}^{(l)}}}$ a constraint check takes place. If its $D\left( {{{\bf{s}}_{u}^{(l)}}} \right) \ge {r^2}$ this node, its children, as well as its not yet visited siblings, are pruned. In addition, in order to avoid redundant calculations which are common to the different minimization problems (and tree searches) of (5), the single-tree-search approach of \cite{SISO_SD} can be employed. According to this, only one tree search takes place but different $r_{\pm ,k}^{2}$ values are used for any of the  minimization problems of (5), with $r_{\pm ,k}^{2}$ being the squared radii related to the two minimization problems (i.e., for $S_{b,i,u}^{\pm 1}$ respectively) of ${{L}_{D}}\left( {{c}_{k}} \right)$ calculation. Whenever a new leaf is reached the $r_{\pm ,k}^{2}$ values of all tree searches to which the resulting symbol vector belongs are updated. For the constraint check at node ${{{\bf{s}}_{u}^{(l)}}}$ the set of tree-searches whose solution can be affected by the corresponding node is identified as $T({{{\bf{s}}_{u}^{(l)}}})$, and pruning is performed if the corresponding PD is larger than all possible  $r_{\pm ,k}^{2}\in T({{{\bf{s}}_{u}^{(l)}}})$,
namely
\begin{equation}
D\left( {{{\bf{s}}_{u}^{(l)}}} \right) > \max_{r_{\pm ,k}^{2}\in T({{{\bf{s}}_{u}^{(l)}}})} r_{\pm ,k}^{2}.
\end{equation}

Minor modifications are needed to this SD in order to take advantage of the proposed approximate information flow. According to those, the proposed SD may perform:

\subsubsection{Selective Soft Information Update (SU)}

As already discussed, the tree searches related to the RWC bits (see (5)) can be skipped. This can be efficiently achieved by zeroing the $r_{\pm ,k}^{2}$ related to the corresponding bits. Then, since the zeroed values are of large $r_{\pm ,k}^{2}$, the constraint of (23) becomes tighter and significant complexity reduction is achieved, as it is also shown in Section IV.

\subsubsection{Performance-Driven Soft Information Clipping (PDC)}

The basic idea behind the proposed performance-driven LLR clipping is to restrict the SD processing by accurately calculating the LLR values only up to the value where the convergence and the required TER are preserved. In this context it would be rational to assume that for the bits which already meet the TER constraint before channel decoding (i.e., $\left| {{L}_{D}}\left( {{c}_{k}} \right) \right|\ge {{\tilde{L}}_{TER}}$) the \emph{average} performance after decoding will be even better. So, reaching the TER before decoding is an indication that further processing may not be required. However, relying only on this assumption to perform LLR clipping is not efficient since this assumption is only valid for the \emph{average} performance and not for each bit. Additionally, performing clipping based on ${L}_{D}$ could (erroneously) result in small $\left| {{{\tilde{L}}}_{A}}\left( {{{\tilde{c}}}_{k}} \right) \right|$ values which would significantly affect the outcome of the channel decoder (see discussion in Section II.B). Therefore, additional consideration should be given to the \emph{extrinsic information} LLR values of the SD (which is the \emph{a-priori} information for the SISO channel decoder) in order to preserve  the system's convergence.

When ${{L}_{E}}\left( {{c}_{k}} \right)$ is of same sign as ${{L}_{D}}\left( {{c}_{k}} \right)$ it is an indication that the iterative process moves towards increasing receiver's confidence on the specific (decoded) bit \cite{tenBrink01}. If, in addition, this bit meets the TER constraint before SISO channel decoding (i.e., $\left| {{L}_{D}}\left( {{c}_{k}} \right) \right|\ge {{\tilde{L}}_{TER}}$) and
the corresponding \emph{extrinsic} information is also strong (i.e., $\left| {{L}_{E}}\left( {{c}_{k}} \right) \right|\ge {{\tilde{L}}_{TER}}$)
it can be roughly assumed that the decoding procedure is mature enough so that the channel decoder will not decrease receiver's confidence for the specific bit in future iterations (i.e., during subsequent iterations will be $sign\left( {{{{L}}}_{A}}\left( {{{{c}}}_{k}} \right) \right)=sign\left( {{{{L}}}_{D}}\left( {{{{c}}}_{k}} \right) \right)=sign\left( {{{{L}}}_{E}}\left( {{{{c}}}_{k}} \right) \right)$). In this case, and since the approximate LLR estimation of the strong LLR values is not expected to significantly affect the outcome of the decoder (see Section II.B) accurate calculation of the LLR values resulting in $\left| {{L}_{E}}\left( {{c}_{k}} \right) \right|=\left| {{{\tilde{L}}}_{A}}\left( {\tilde{c}_{k}}={{\pi }^{-1}}\left( {{c}_{k}} \right) \right) \right|>{{\tilde{L}}_{TER}}$ is not required. On the contrary, if the $sign\left\{ {{L}_{E}}\left( {{c}_{k}} \right) \right\}\ne sign\left\{ {{L}_{D}}\left( {{c}_{k}} \right) \right\}$ (i.e., the receiver's confidence for the candidate decoded bit is not increasing) it is an indication that the iterative decoding process is not yet mature, so LLR clipping should be avoided in order to preserve the convergence properties. It is significant to note that if LLR clipping is (erroneously) performed on a bit which converges opposite to the finally decoded bit (i.e., $sign\left( {{{{L}}}_{E}}\left( {{{{c}}}_{k}} \right) \right)=sign\left( {{{{L}}}_{D}}\left( {{{{c}}}_{k}} \right) \right)\ne {{\hat{c}}_{k,final}}$) it is not expected to negatively affect the performance since clipping practically bounds the effects of this erroneous convergence. This is also verified in Section IV.
% NOTE: if clipping is performed in other bits, with sign L_e different than L_d then clipping
% prevents from good convergence since it does not leave system converge to the right

According to the previous discussion, the SD search hypersphere should be reduced in a way that both the convergence and the TER performance after the SD (and before the channel decoder) are preserved. Equivalently, LLR approximation is allowed only when both the ${{L}_{E}}\left( {{c}_{k}} \right)$ and the ${{L}_{D}}\left( {{c}_{k}} \right)$ values are larger than ${\tilde{L}}_{TER}$. As already discussed, LLR clipping approaches which account only for the performance  before decoding can result in small $\left| {{{\tilde{L}}}_{A}}\left( {{{\tilde{c}}}_{k}} \right) \right|$ values which would consequently affect the outcome of the channel decoder and therefore the resulting performance. In addition, as it will be show in the sequel, clipping approaches targeting only $\left| {{L}_{E}}\left( {{c}_{k}} \right) \right|\le {{\tilde{L}}_{TER}}$ result in performance degradation.

From (14), it follows that
\begin{equation}
{{L}_{D}}\left( {{c}_{k}} \right)=\left\{ \begin{matrix}
   \lambda _{k}^{\overline{MAP}}-\lambda _{{}}^{MAP},~c_{k}^{MAP}=sign\left\{ {{L}_{D}}\left( {{c}_{k}} \right) \right\}=+1  \\
   \lambda _{{}}^{MAP}-\lambda _{k}^{\overline{MAP}},~c_{k}^{MAP}=sign\left\{ {{L}_{D}}\left( {{c}_{k}} \right) \right\}=-1  \\
\end{matrix} \right.
\end{equation}
with
$
{\lambda }^{MAP} = \mathop {\min }\limits_{{\bf{s}}_{u} \in S} \left\{ {I({\bf{s}}_{u})} \right\}
$ being the minimum ${{I}}\left( {{{\bf{s}}_{u}}} \right)$ value found during the corresponding unconstraint single-tree search, $c_{k}^{MAP}$ the $k$-th bit value of the symbol vector providing ${\lambda }^{MAP}$, and
$
\lambda _{k}^{\overline{MAP}}= \mathop {\min }\limits_{{\bf{s}}_{u} \in S_{k}^{ \overline{MAP}}} \left\{ {I({\bf{s}}_{u})} \right\}\
$ where $S_{k}^{ \overline{MAP}}$ are the sub-sets of possible ${{\bf{s}}_u}$ symbols sequences having their $k$-th bit value opposite to the one of the MAP solution. Therefore, the search space for any of the trees can be reduced to a hypersphere of
%\[r_{PDC,\pm ,k }^{2}={{\hat{\lambda }}^{MAP}}+\left| {{L}_{A}}\left( {{c}_{k}} \right) \right|+{{\tilde{L}}_{TER}}+\]
\begin{equation}
r_{PDC,\pm ,k }^{2}={{\hat{\lambda }}^{MAP}}+\left| {{L}_{A}}\left( {{c}_{k}} \right) \right|+{{\tilde{L}}_{TER}}+
\frac{1}{2}\left( \hat{c}_{k}^{MAP}-sign\left\{ {{L}_{A}}\left( {{c}_{k}} \right) \right\} \right){{L}_{A}}\left( {{c}_{k}} \right).
\end{equation}
where  ${{{\hat{\lambda }}}^{MAP}}$ is the minimum ${{I}}\left( {{{\bf{s}}_{u}}} \right)$ value already found and $\hat{c}_{k}^{MAP}$ is the related $k$-th bit value.
Then, the corresponding pruning constraint can become
\begin{equation}
D\left( {{{\bf{s}}_{u}^{(l)}}} \right) > \max_{r_{\pm ,k}^{2}\in T({{{\bf{s}}_{u}^{(l)}}})} \min{\left\{ r_{\pm ,k}^{2}, r_{PDC, \pm ,k}^{2} \right\}}.
\end{equation}
Any time a new candidate ${{{\hat{\lambda }}}^{MAP}}$ is found the corresponding $r_{\pm ,k}^{2}$ values can be updated to $r_{\pm ,k}^{2}\leftarrow \max \{r_{\pm ,k}^{2},r_{PDC,\pm ,k}^{2}\}$. This is preformed in order to produce a clipped LLR value even if no solution of (5) lies in the search hypershere for the corresponfing bit. Both selective LLR update and hypersphere reduction result in a tighter constraint check than the typical (see (20)) and thus, in reduced SD processing.

This search space reduction results in bounded $\left| {{L}_{D}}\left( {{c}_{k}} \right) \right|$ values. From (21) and (22) it can be easily deduced that for ${{{\hat{\lambda }}}^{MAP}}={{{{\lambda }}}^{MAP}}$ the corresponding $L_{E}^{{{\left| {{L}_{D}}\left( {{c}_{k}} \right) \right|}_{\max }}}\left( {{c}_{k}} \right)$ values which maximize $\left| {{L}_{D}}\left( {{c}_{k}} \right) \right|$, are
%\begin{equation}
%L_{E}^{{{\left| {{L}_{D}}\left( {{c}_{k}} \right) \right|}_{\max }}}\left( {{c}_{k}} \right)=\left\{ \begin{matrix}
 %  sign\left\{ {{L}_{D}}\left( {{c}_{k}} \right) \right\}{{{\tilde{L}}}_{TER}},  \\
 % for~sign\left\{ {{L}_{A}}\left( {{c}_{k}} \right) \right\}=sign\left\{ {{L}_{D}}\left( {{c}_{k}} \right) \right\}  \\
%  \\
%  sign\left\{ {{L}_{D}}\left( {{c}_{k}} \right) \right\}\left( {{{\tilde{L}}}_{TER}}+\left| {{L}_{A}}\left( {{c}_{k}} \right) \right| \right),  \\
%   otherwise  \\
%\end{matrix} \right.
%\end{equation}
\begin{equation}
L_{E}^{{{\left| {{L}_{D}}\left( {{c}_{k}} \right) \right|}_{\max }}}\left( {{c}_{k}} \right)=\left\{ \begin{matrix}
   sign\left\{ {{L}_{D}}\left( {{c}_{k}} \right) \right\}{{{\tilde{L}}}_{TER}}, & sign\left\{ {{L}_{A}}\left( {{c}_{k}} \right) \right\}=sign\left\{ {{L}_{D}}\left( {{c}_{k}} \right) \right\}  \\
   sign\left\{ {{L}_{D}}\left( {{c}_{k}} \right) \right\}\left( {{{\tilde{L}}}_{TER}}+\left| {{L}_{A}}\left( {{c}_{k}} \right) \right| \right), & else  \\
\end{matrix} \right..
\end{equation}
From the above equation it becomes apparent that when the \emph{extrinsic} information of the SD is of the same sign as its \emph{a-posteriori} information the clipping value is such that no processing is spent for calculating values which exceed the TER constraint. For example if $sign\left\{ {{L}_{A}}\left( {{c}_{k}} \right) \right\}=sign\left\{ {{L}_{D}}\left( {{c}_{k}} \right) \right\}=1$, the maximum ${{L}_{E}}\left( {{c}_{k}} \right)$ value equals $\tilde{L}_{TER}$. In addition, the proposed clipping preserves the ability of the bits to reach the TER before decoding. For example if
$sign\left\{ {{L}_{A}}\left( {{c}_{k}} \right) \right\} \ne sign\left\{ {{L}_{D}}\left( {{c}_{k}} \right) \right\}=1$ then $L_{D}^{{{\left| {{L}_{D}}\left( {{c}_{k}} \right) \right|}_{\max }}}\left( {{c}_{k}} \right)={{L}_{A}}\left( {{c}_{k}} \right)+L_{E}^{{{\left| {{L}_{D}}\left( {{c}_{k}} \right) \right|}_{\max }}}\left( {{c}_{k}} \right)={{L}_{A}}\left( {{c}_{k}} \right)+\left| {{L}_{A}}\left( {{c}_{k}} \right) \right|+{{\tilde{L}}_{TER}}={{\tilde{L}}_{TER}}$.

The fact that the clipping process employs the $\hat{c}_{k}^{MAP}$ estimates instead of the exact ${c}_{k}^{MAP}$ may sometimes lead to tighter LLR clipping than wanted. However, in Section IV it is shown that this does not have any considerable effect in scheme's performance.

\subsubsection{Simplified Performance-Driven Soft Information Clipping (sPDC)}

A simplified PD-PDC can be acquired when it is not of interest to preserve the TER performance before channel decoding, namely, when clipping is allowed for bits with $\left| {{L}_{D}}\left( {{c}_{k}} \right) \right| < {{\tilde{L}}_{TER}}$  . Then, the search hypersphere can be reduced to

%\[r_{sPDC,\pm ,k }^{2}={{\hat{\lambda }}^{MAP}}+\left| {{L}_{A}}\left( {{c}_{k}} \right) \right|+{{\tilde{L}}_{TER}}+\]
\begin{equation}
r_{sPDC,\pm ,k }^{2}={{\hat{\lambda }}^{MAP}}+\left| {{L}_{A}}\left( {{c}_{k}} \right) \right|+{{\tilde{L}}_{TER}}+
\left( \hat{c}_{k}^{MAP}-sign\left\{ {{L}_{A}}\left( {{c}_{k}} \right) \right\} \right){{L}_{A}}\left( {{c}_{k}} \right).
\end{equation}
Equivalently to (24), $L_{E}^{{{\left| {{L}_{D}}\left( {{c}_{k}} \right) \right|}_{\max }}}\left( {{c}_{k}} \right)=sign\left\{ {{L}_{D}}\left( {{c}_{k}} \right) \right\}{{\tilde{L}}_{TER}}$ even for $sign\left\{ {{L}_{A}}\left( {{c}_{k}} \right) \right\} \ne sign\left\{ {{L}_{D}}\left( {{c}_{k}} \right) \right\}$. Therefore, for the previous example $L_{D}^{{{\left| {{L}_{D}}\left( {{c}_{k}} \right) \right|}_{\max }}}\left( {{c}_{k}} \right)={{L}_{A}}\left( {{c}_{k}} \right)+{{\tilde{L}}_{TER}}={{\tilde{L}}_{TER}}-\left| {{L}_{A}}\left( {{c}_{k}} \right) \right|\le {{\tilde{L}}_{TER}}$. However, as it is shown in Section IV, not preserving the TER before channel decoding results in a noticeable performance degradation without providing any significant complexity gain.

\subsubsection{Decoder-Aware Performance-Driven Soft Information Clipping (DA-PDC)}

Tighter LLR clipping than the one of the PDC can be performed by making some further (approximate) assumptions on the ``expected'' reliability (i.e., LLR amplitude) increase provided by the SISO channel decoder. In detail, if after the SD processing the sign of the demapped bit is sustained (i.e., $sign\left( {{{{L}}}_{A}}\left( {{{{c}}}_{k}} \right) \right)=sign\left( {{{{L}}}_{D}}\left( {{{{c}}}_{k}} \right) \right)$) it means that the iterative process increases its confidence for this (hard) decoded bit. Then, it is approximately assumed that the sign of the decoder's \emph{extrinsic} information (which will be the SD's \emph{a-priori} information) will remain constant, and the magnitude will be at least the same. This can be typically observed when the iterative process is close to the state of constant information flow where the most significantly contributing sequences in (15) remain the same and the related \emph{a-priori} information has already reached its constant flow state or still increases \cite{tenBrink01}.
Under this assumptions, the search hypersphere can be further tightened to accurately calculate only the ${{L}_{E}}\left( {{c}_{k}} \right)$ values of those bits which cannot reach the TER performance even after SISO channel decoding. In detail, the search space can be reduced to a hypersphere of
%\[r_{DA-PDC,\pm ,k}^{2}={{\hat{\lambda }}^{MAP}}+{{\tilde{L}}_{TER}}+\]
\begin{equation}
r_{DA-PDC,\pm ,k}^{2}={{\hat{\lambda }}^{MAP}}+{{\tilde{L}}_{TER}}+
\frac{1}{2}\left( \hat{c}_{k}^{MAP}-sign\left\{ {{L}_{A}}\left( {{c}_{k}} \right) \right\} \right)\left( \hat{c}_{k}^{MAP}\left| {{L}_{A}}\left( {{c}_{k}} \right) \right|+{{L}_{A}}\left( {{c}_{k}} \right) \right)
\end{equation}
with the corresponding constraint to become
\begin{equation}
D\left( {{{\bf{s}}_{u}^{(l)}}} \right) > \max_{r_{\pm ,k}^{2}\in T({{{\bf{s}}_{u}^{(l)}}})} \min{\left\{ r_{\pm ,k}^{2}, r_{DA-PDC, \pm ,k}^{2} \right\}}.
\end{equation}
Then, it can be easily shown that the  $L_{E}^{{{\left| {{L}_{D}}\left( {{c}_{k}} \right) \right|}_{\max }}}\left( {{c}_{k}} \right)$ values which maximize the $\left| {{L}_{D}}\left( {{c}_{k}} \right) \right|$ value, are
%\begin{equation}
%L_{E}^{{{\left| {{L}_{D}}\left( {{c}_{k}} \right) \right|}_{\max }}}\left( {{c}_{k}} \right)=\left\{ \begin{matrix}
%   sign\left\{ {{L}_{D}}\left( {{c}_{k}} \right) \right\}{{{\tilde{L}}}_{TER}}-{{L}_{A}}\left( {{c}_{k}} \right),  \\
%   for~sign\left\{ {{L}_{A}}\left( {{c}_{k}} \right) \right\}=sign\left\{ {{L}_{D}}\left( {{c}_{k}} \right) \right\}  \\
%   \\
%   sign\left\{ {{L}_{D}}\left( {{c}_{k}} \right) \right\}\left( {{{\tilde{L}}}_{TER}}+\left| {{L}_{A}}\left( {{c}_{k}} \right) \right| \right),  \\
%   otherwise  \\
%\end{matrix} \right.
%\end{equation}
\begin{equation}
L_{E}^{{{\left| {{L}_{D}}\left( {{c}_{k}} \right) \right|}_{\max }}}\left( {{c}_{k}} \right)=\left\{ \begin{matrix}
   sign\left\{ {{L}_{D}}\left( {{c}_{k}} \right) \right\}{{{\tilde{L}}}_{TER}}-{{L}_{A}}\left( {{c}_{k}} \right), & sign\left\{ {{L}_{A}}\left( {{c}_{k}} \right) \right\}=sign\left\{ {{L}_{D}}\left( {{c}_{k}} \right) \right\}  \\
   sign\left\{ {{L}_{D}}\left( {{c}_{k}} \right) \right\}\left( {{{\tilde{L}}}_{TER}}+\left| {{L}_{A}}\left( {{c}_{k}} \right) \right| \right), & else  \\
\end{matrix} \right.
\end{equation}
From (28) it becomes apparent that with such a hypersphere reduction, when $sign\left\{ {{L}_{A}}\left( {{c}_{k}} \right) \right\}=sign\left\{ {{L}_{D}}\left( {{c}_{k}} \right) \right\}$, only values which are not expected to reach TER after decoding are accurately calculated while, if $sign\left\{ {{L}_{A}}\left( {{c}_{k}} \right) \right\} \ne sign\left\{ {{L}_{D}}\left( {{c}_{k}} \right) \right\}$, extrinsic information values up to ${{\tilde{L}}_{TER}}$ are accurately calculated  similarly to the PIDC. For example if $sign\left\{ {{L}_{A}}\left( {{c}_{k}} \right) \right\}=sign\left\{ {{L}_{D}}\left( {{c}_{k}} \right) \right\}=1$ the corresponding maximum soft information input to the SISO channel decoder for the bit $\tilde{c}_{k}={{\pi }^{-1}}\left( {{c}_{k}} \right)$ will be $\tilde{{L}}_{A}\left({\tilde{c}_{k}} \right) ={{\tilde{L}}_{TER}}-{{L}_{A}}\left({{c}_{k}} \right)={{\tilde{L}}_{TER}}- {\tilde{L}_{E}}\left({\tilde{c}_{k}} \right)$. Therefore, if the $\left| {{{\tilde{L}}}_{E}}\left( {{{\tilde{c}}}_{k}} \right) \right|$ of the current iteration is at least equal to the previous, $\left| {{{\tilde{L}}}_{D}}\left( {{{\tilde{c}}}_{k}} \right) \right|$ will meet the TER requirement after decoding.

For the bits whose LLR value is of such a high magnitude that a solution of (5) does not lie in this shrunken hypershere, clipping is performed according to the PIDC by updating $r_{\pm ,k}^{2}$ as $r_{\pm ,k}^{2}\leftarrow \max \{r_{\pm ,k}^{2},r_{PDC,\pm ,k}^{2}\}$ any time a new candidate ${{{\hat{\lambda }}}^{MAP}}$ is found.
Then, if the \emph{a-posteriori} information of the bit does not belong in the shrunken hypersphere its LLR value will be set to such a value that its \emph{extrinsic} information reaches ${{\tilde{L}}_{TER}}$, similarly to PIDC.

According to this last approximation, a bit with loose \emph{extrinsic} and strong \emph{a-priori} information (and $sign\left\{ {{L}_{A}}\left( {{c}_{k}} \right) \right\}=sign\left\{ {{L}_{D}}\left( {{c}_{k}} \right) \right\}$) may be erroneously assumed to have reached the TER. However, if this wrong assumption is critical for the convergence of its neighboring $\Lambda$  bits, it will be  manifested as a more loose \emph{extrinsic} information at the channel decoder output (or more loose SD \emph{a-priori} information) over the next iteration, similarly to what has been discussed in Section III.B. Subsequently, this will result in an increase of the search hypersphere during the next iteration and, thus, in more accurate LLR estimation.

%In addition, if, according to the first assumption, the gain of the channel decoder is overestimated this is just expected to be manifested as a negative effect on convergence rate.
\subsubsection{Simplified Decoder-Aware Performance-Driven Soft Information Clipping (sDA-PDC)}

Similarly to the sPDC, when the prevention of the TER performance before decoding is not targeted, the hypersphere can be reduced to $r_{sDA-PDC,\pm ,k}^{2}={{\hat{\lambda }}^{MAP}}+{{\tilde{L}}_{TER}}$, resulting in $L_{E}^{{{\left| {{L}_{D}}\left( {{c}_{k}} \right) \right|}_{\max }}}\left( {{c}_{k}} \right)=sign\left\{ {{L}_{D}}\left( {{c}_{k}} \right) \right\}{{\tilde{L}}_{TER}}-{{L}_{A}}\left( {{c}_{k}} \right)$. In such a case the $r_{\pm ,k}^{2}\leftarrow \max \{r_{\pm ,k}^{2},r_{sPDC,\pm ,k}^{2}\}$ update is performed any time a new candidate ${{{\hat{\lambda }}}^{MAP}}$ is found. Similarly to the sPDC, and as shown in Section V, this approach does not provide any significant complexity gain compared to the DA-PDC but it results in a noticeable performance degradation.

The discussed LLR clipping approaches are selected in a way that the necessity of accurately calculating both the $\left| {{L}_{E}} \right|$ and $\left| {{L}_{D}} \right|$ values up to $\tilde{L}_{TER}$ is revealed. However, the proposed manifestations are not unique and several alternatives of similar complexity can be found, which still meet the same criteria but in a less tight way. For example, it can be easily verified that similarly to the PDC, a reduced hypersphere of $r_{\pm ,k}^{2}={{\hat{\lambda }}^{MAP}}+\left| {{L}_{A}}\left( {{c}_{k}} \right) \right|+{{\tilde{L}}_{TER}}$ would result in clipped values only after both ${{L}_{D}}\left( {{c}_{k}} \right)$ and ${{L}_{E}}\left( {{c}_{k}} \right)$  meet the TER, but it would allow a larger $L_{E}^{{{\left| {{L}_{D}}\left( {{c}_{k}} \right) \right|}_{\max }}}\left( {{c}_{k}} \right)$
when $sign\left\{ {{L}_{A}}\left( {{c}_{k}} \right) \right\}\ne sign\left\{ {{L}_{D}}\left( {{c}_{k}} \right) \right\}$. This, can be shown by simulations, to result only in an incremental increase in the number of visited nodes.

% to slight increase einai xwris SU. me SU exw small increase se low iterations kai small gain in high iteration (giati me  megalytero clipping value, ftanw pio grigora to RWC).

\subsection{Scenario-Adaptive SISO Channel Decoder}

 As already discussed in Section III.A, step 3, the proposed scenario-adaptive SISO channel decoder performs decoding only on a subset of LLR values. Typical SISO channel decoder realizations operate in the \emph{log} domain and employ the $max^{*}$ function in order to replace the computationally expensive multiplications with additions as described in \cite{BenedettoII}. Then, as it is shown in \cite{Schurgers99,Schurgers01}, the most expensive operations become the necessary, energy consuming, memory accesses and especially the ones related to the state metric storages. The significance of reducing those memory accesses is emphasized in \cite{Schurgers99} where additional processing and register file storage is paid for this reason. However, even with such approaches, the number of accesses cannot be substantially reduced due to the energy overhead of the processing and the register file storage. In the sequel, equivalently to \cite{Dan10}, it is discussed how the selective LLR update of the non-RWC bits may result in reduced number of state metric storages. However, it is significant to note that this discussion is just indicative since the selective updates cannot be quantified into energy savings without considering a specific implementation, which is beyond the scope of this work.

 For a convolutional code of 1/2 rate and with ${{\tilde{c}}_{x,t}}(e)$ the encoder output bits for a transition $e$ from the state $s$ to $s'$ at coding time $t$ (with $s$,$s'$=$0,..,N_s-1$ and  $x=0,1$) the corresponding ${{\tilde{L}}_{D}}\left( {{{\tilde{c}}}_{x,t}} \right)$ can be expressed as \cite{BenedettoII}
\begin{equation}
{{\tilde{L}}_{D}}\left( {{{\tilde{c}}}_{x,t}} \right)={{\underset{e:{{{\tilde{c}}}_f{x,t}}=1}{\mathop{\max }}\,}^{*}}\left[ {{\delta }_{t}}(e) \right]-{{\underset{e:{{{\tilde{c}}}_{x,t}}=-1}{\mathop{\max }}\,}^{*}}\left[ {{\delta }_{t}}(e) \right]
\end{equation}
with
\begin{equation}
{{\delta }_{t}}(e)={{\alpha }_{t-1}}[s]+{\tilde{c}_{0,t}}(e){{{\tilde{L}}}_{A}}\left( {{{\tilde{c}}}_{0,t}}(e) \right)+{\tilde{c}_{1,t}}(e){{{\tilde{L}}}_{A}}\left( {{{\tilde{c}}}_{1,t}}(e) \right)+{{\beta }_{t}}[s']
\end{equation}
and ${{\alpha}_{t}}$, ${{\beta}_{t}}$ being the state metrics obtained through the following forward and backward recursions
\begin{equation}
{{\alpha }_{t}}(w)={{\underset{e:s'=w}{\mathop{\max }}\,}^{*}}\left[ {{\alpha }_{t-1}}(s)+{\tilde{c}_{0,t}}(e){{{\tilde{L}}}_{A}}\left( {{{\tilde{c}}}_{0,t}}(e) \right)+{\tilde{c}_{1,t}}(e){{{\tilde{L}}}_{A}}\left( {{{\tilde{c}}}_{1,t}}(e) \right) \right]
\end{equation}
\begin{equation}
{{\beta }_{t}}(w)={{\underset{e:s=w}{\mathop{\max }}\,}^{*}}\left[ {{\beta }_{t+1}}(s')+{\tilde{c}_{0,t+1}}(e){{{\tilde{L}}}_{A}}\left( {{{\tilde{c}}}_{0,t+1}}(e) \right)+{\tilde{c}_{1,t+1}}(e){{{\tilde{L}}}_{A}}\left( {{{\tilde{c}}}_{1,t+1}}(e) \right) \right].
\end{equation}

As discussed in \cite{Schurgers01}, the ${\alpha}_{t}(w)$ values can be calculated and overwritten immediately as they are not required in future calculations. On the other hand, typically, all ${\beta}_{t}(w)$ metrics need to be stored. However, for selective (per bit) channel decoding only the subset of ${\beta}_{t}(w)$ values related to the decoded bits needs to be stored, resulting in potential energy consumption savings.

\subsection{Complexity Issues}
In this subsection, some complexity issues are discussed without considering the early-stopping control, since it is not required from the proposed scheme and as it can be replaced by other similar early-stopping approaches.

An additional memory of $K$ bits is required for storing the flag sequence $G^{(q)}\left( {{k}} \right)$. The additional required interleaving effort for the flag sequence is a small portion of the overall interleaving one, since the introduced one bit overhead is typically small compared to the number of bits employed to represent the extrinsic LLR values in fixed point arithmetics. De-interleaving is not required since the flag sequence does not change within iterations. An additional complexity increase of $2K$ real number comparisons is introduced for RWC bit identification. The one-bit comparisons which are required to identify the position of the RWC bits at the SD and SISO outer decoder side are assumed negligible compared to the real ones.

In order to assess the SD complexity gains via simulations in Section IV the number of the visited nodes is employed as an indicative measure. However, similar results hold for other measures as the number of expanded nodes or the number of required partial distance calculations.

As already discussed, the energy savings at the SISO channel decoder side can be quantified only for specific implementations. However, since the main potential gain is expected to originate from the minimization of the memory accesses, two measures are indicatively considered related to the different types of memory accesses. The first one is the number of the non-RCW bits which is related to the number of accesses required for the channel decoder's \emph{a-priori} information update. The other is the number of the required  $\beta={{\left[ {{\beta }_{t}}(0),...,{{\beta }_{t}}({{N}_{s}}-1) \right]}^{T}}$ calculations which is related to the state metric storages.

\section{Simulations}
A $4\times4$ MIMO system is assumed operating over a spatially and temporally uncorrelated Rayleigh flat-fading channel. The encoded bits are mapped onto 16-QAM via Gray coding. A systematic  ${(5/7)_8}$
 recursive convolutional code of rate 1/2 is employed with code block of $18432$ bits. The log-MAP BCJR algorithm has been employed for SISO channel decoding. Early stopping control of error rate equal to the TER is always assumed (even with the typical SD).

 In Fig. 2 the BER performance of the proposed scheme is depicted for an SNR of 7 dB. Three proposed SD approaches are compared to the typical SD. These are the SU, the SU \& PDC, and the SU \& DA-PDC. The TER is set to $2\cdot10^{-3}$, slightly lower than the best achievable BER ($\approx 2.2\cdot10^{-3}$). Selective SISO channel decoding is employed with $w=1$. It is shown that the proposed RWC identification methodology is that reliable where negligible performance loss is observed even for selective SISO channel decoding of minimum $w=1$. In Fig. 3 the (cumulative over iterations) complexity of the corresponding SD approaches is shown, while in Fig. 4 the (cumulative) required $\beta$ stores and the number of the non-RWC bits are depicted. It is shown that, at iteration 5, the SU provides an SD complexity gain of about 28\%. The SU \& PDC approach provides a complexity gain of about 71\% compared to the SU approach, while the SU \& DA-PDC SD provides an additional complexity gain of about 25\% compared to the SU \& PDC one, with the total complexity gain, compared to the typical, reaching 84\%.
 At the same time, for the SU \& DA-PDC soft demapper, the gain related to the $\beta$ stores and the number of the non-RWC bits reaches the 41\% and the 46\% accordingly. It is also shown that the corresponding store requirements are slightly dependent on the SD approach and only at high iterations. This is an expected behavior since a ``good" soft-information approximation should not affect the process towards bit convergence, but only the point where the convergence stops.

 In Figs. 5 and 6 the efficiency of the proposed performance-driven clipping methods is depicted in terms of BER performance and SD complexity savings, when combined with SU. An SNR of 7 dB is assumed with a TER of $2\cdot10^{-3}$ and full SISO channel decoding. It is shown that both the PDC and DA-PDC approaches allow reaching the TER performance with a negligible performance loss (which is visible only for the DA-PDC). On the contrary, their simplified versions result in non-negligible performance loss despite their slightly increased complexity (i.e., tighter clipping can result in delayed over iterations RWC bit identification). As shown, at iteration 5, the (s)DA-PDC approaches can provide an additional complexity gain of about 25\% compared to the (s)PDC ones.

 In Figs. 7 and 8 the proposed approach is compared to the performance of a typical iterative scheme for the SNRs of 7 and 8 dB respectively and for several TER values. An SD with SU \& DA-PDC and selective channel decoding of $w=1$ are employed. As shown in Figs. 9 and 10, at iteration 3 for example, a TER reduction of an an order of magnitude results in SD complexity gains of 30-42\%. In addition, the overall SD complexity gain ranges from 82 to 90\%.
 %in7 dB case, an order of magnitude TER reduction results in SD complexity reduction of about 42\% at iteration 3, while for $TER=2\cdot10^{-2}$ the corresponding gain compared to the typical case is about 85\% For the 9 db case at iteration 3 one order of magnitude reduction gives gain 30-37\% and compared to the typical 82-90.
 In Figs. 11, 12 it is shown that significant gains in the number of memory accesses can be achieved only over higher iterations where the number of RWC bits is adequately high. For example, for 9 dB and iteration 3, gains of 33\% and 38\% are observed in the number of the $\beta$ stores and the number of the non-RWC bits respectively, for a TER of $10^{-4}$, while for a TER of $10^{-2}$ the convergence process stops earlier and the corresponding gains become 21\% and 26\%.

\section{Conclusion}
An iterative receiver processing framework of approximate soft information exchange has been proposed which allows the adjustment of the receiver processing requirements (i.e., of the soft-output detector and of the SISO channel decoder) to the transmission conditions and the required BER. In this context, several \emph{performance-driven} LLR clipping methods together with partial soft information update are proposed in order to adjust the complexity of the receiver to the target performance. Despite the small additional overhead the approach can provide substantial complexity savings both at the soft-output detector and the channel decoder.

\linespread{1.5}
% or


%\appendices

%\section{}

%\section*{Acknowledgment}

\ifCLASSOPTIONcaptionsoff
  \newpage
\fi



%\bibliographystyle{IEEEtran}
%%\begin{thebibliography}{1}

%%\bibitem{IEEEhowto:kopka}
%%H.~Kopka and P.~W. Daly, \emph{A Guide to \LaTeX}, 3rd~ed.\hskip 1em plus
%%  0.5em minus 0.4em\relax Harlow, England: Addison-Wesley, 1999.
%\newpage
%%\end{thebibliography}

%\newpage
\bibliographystyle{IEEEtran}
\bibliography{JustNeededSD_Bib}
%\end{spacing}

%\newpage
%~
%\newpage

 \begin{figure}[d]
  \centering
\includegraphics[width=0.8\linewidth]{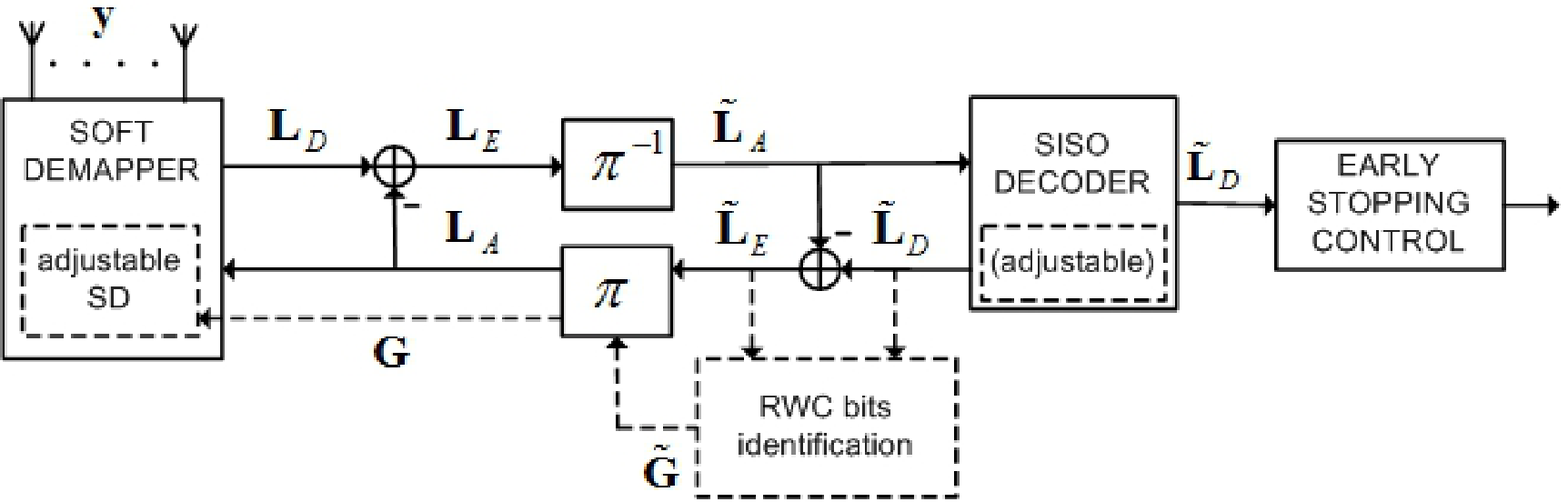}\\
  \caption{Block diagram of a typical iterative scheme (solid lines) with the proposed modification (dashed lines) for approximate soft information flow.}
\end{figure}

\begin{figure}
  \centering
\includegraphics[width=0.7\linewidth]{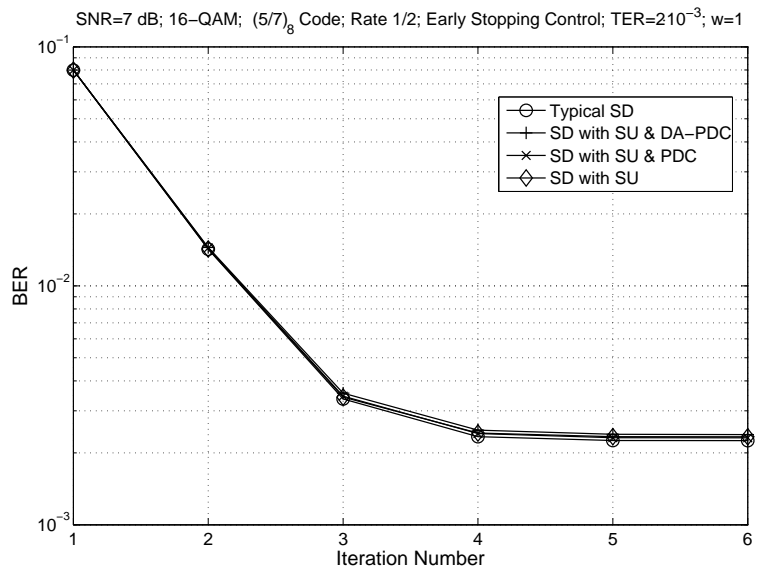}\\
  \caption{BER performance for a system with selective channel decoding of $w=1$ and soft demapping with SU, SU \& PDC, and SU \& DA-PDC at 7dB.}
\end{figure}

\begin{figure}
  \centering
\includegraphics[width=0.7\linewidth]{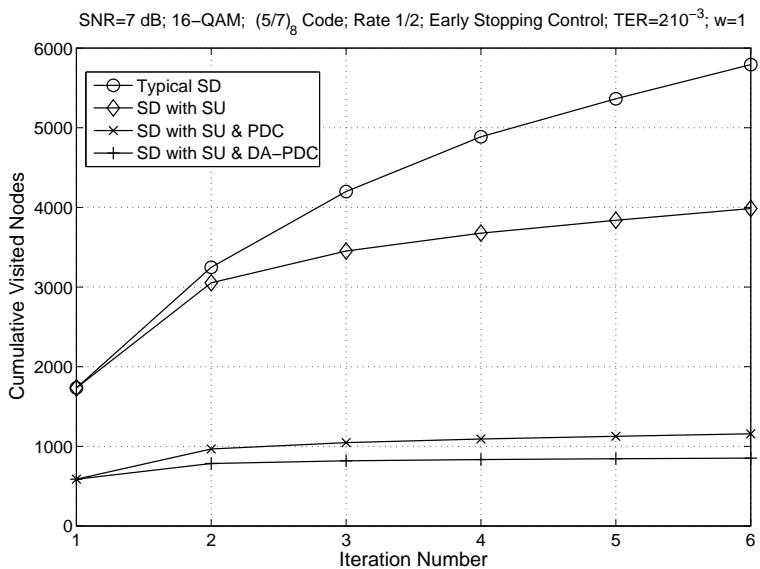}\\
  \caption{Soft demapping complexity for a system with selective channel decoding of $w=1$ and soft demapping with SU, SU \& PDC, and SU \& DA-PDC at 7dB.}
\end{figure}

\newpage
\begin{figure}
  \centering
\includegraphics[width=0.7\linewidth]{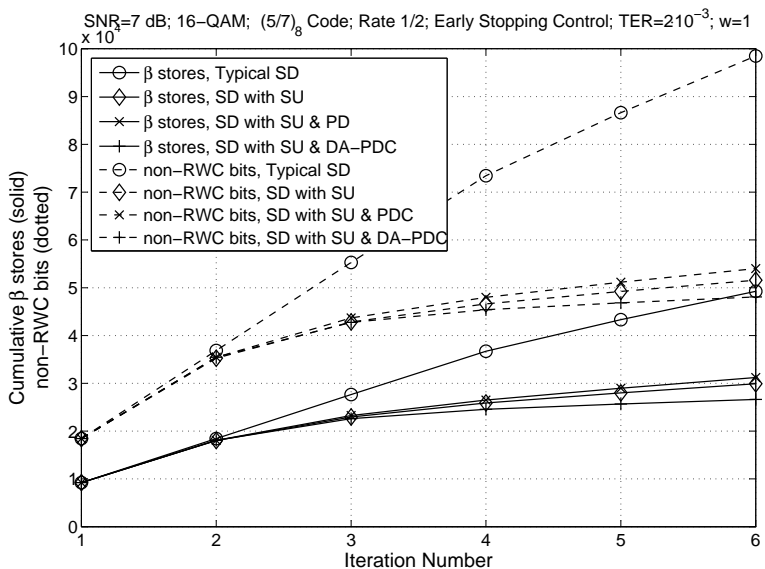}\\
  \caption{Memory store requirements for a system with selective channel decoding of $w=1$ and soft demapping with SU, SU \& PDC, and SU \& DA-PDC at 7dB.}
\end{figure}

\newpage
\begin{figure}
  \centering
\includegraphics[width=0.7\linewidth]{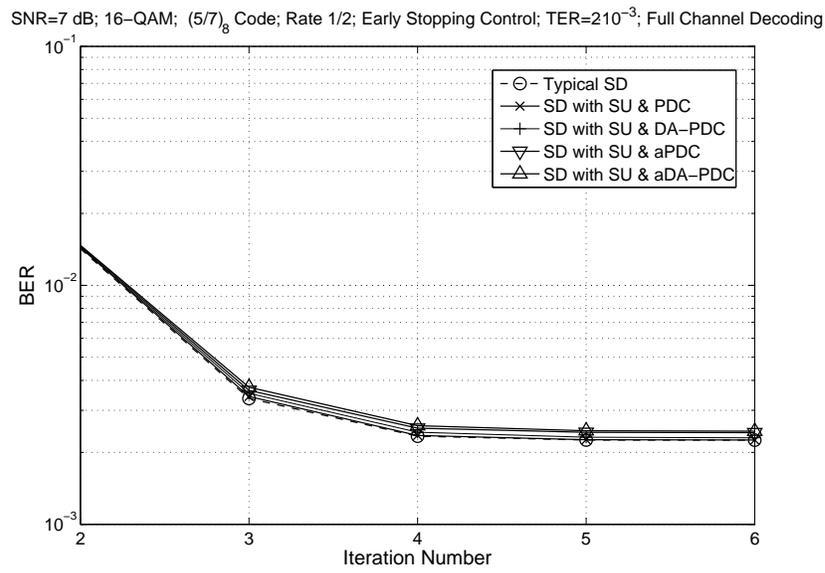}\\
  \caption{BER performance for a system with full channel decoding and soft demappers with SU and different clipping approaches at 7 dB.}
\end{figure}

\newpage
\begin{figure}
  \centering
\includegraphics[width=0.7\linewidth]{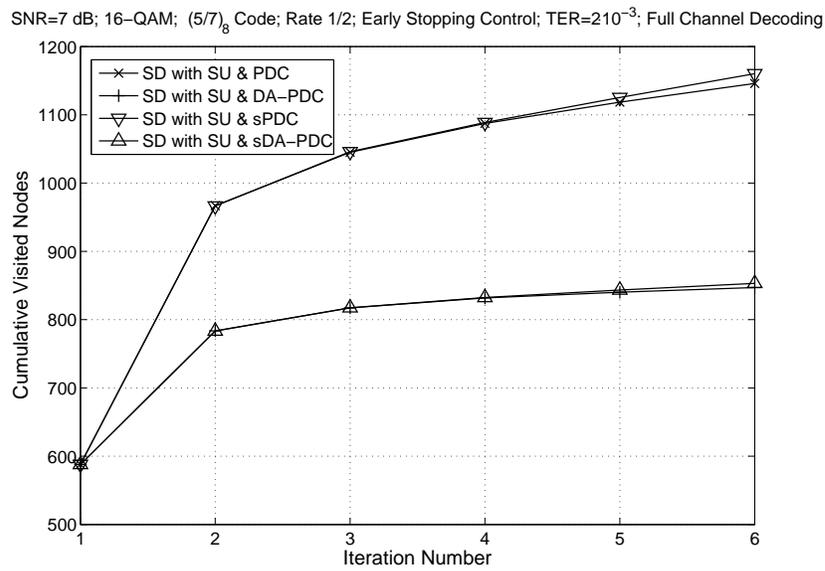}\\
  \caption{Soft demapping complexity for a system with full channel decoding and soft demappers with SU and different clipping approaches at 7 dB.}
\end{figure}

\newpage
\begin{figure}
  \centering
\includegraphics[width=0.7\linewidth]{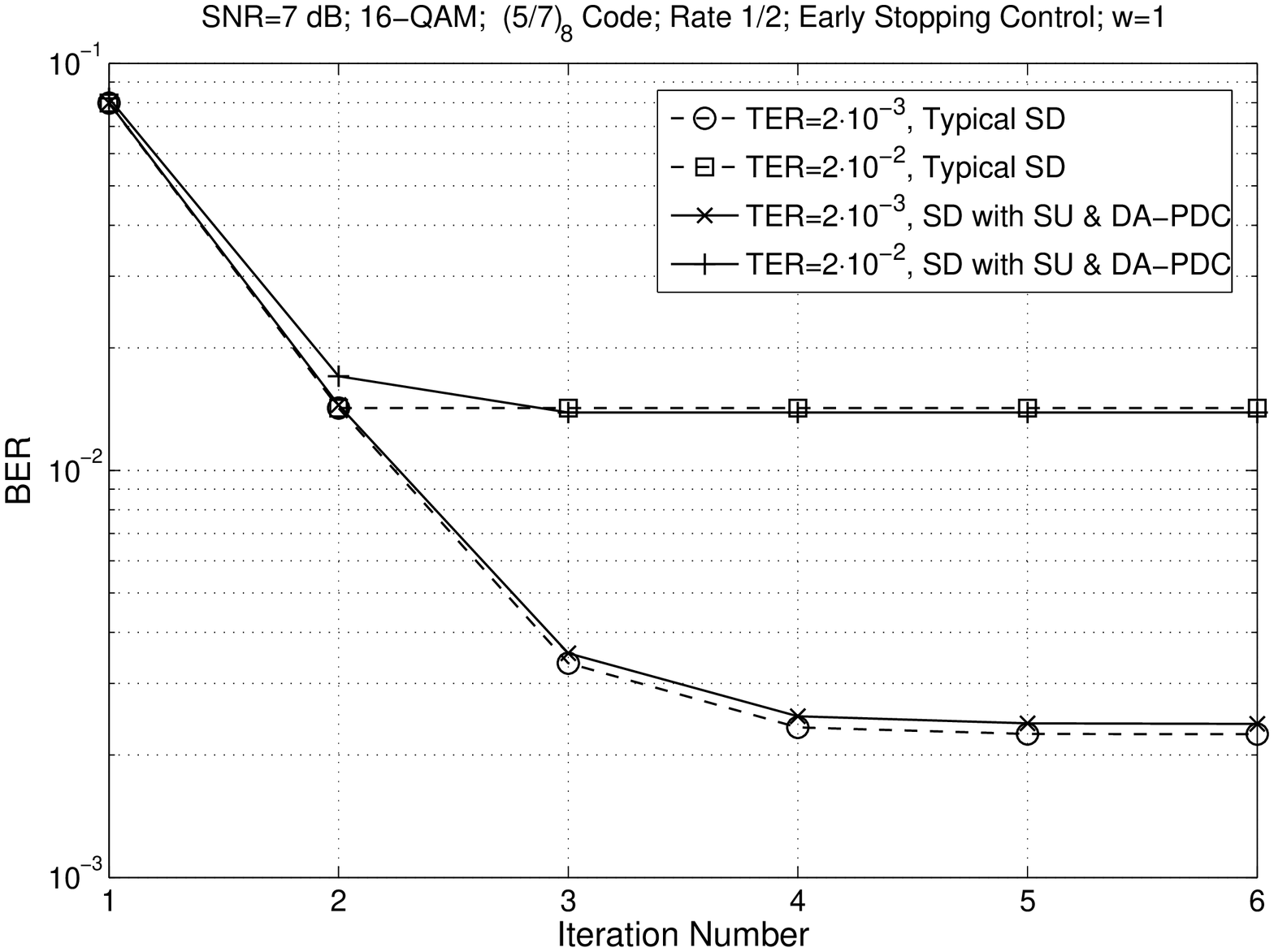}\\
  \caption{BER performance for a system with selective channel decoding of $w=1$,  SU \& DA-PDC soft demapping and several TER values at 7 dB.}
\end{figure}

\newpage
\begin{figure}
  \centering
\includegraphics[width=0.7\linewidth]{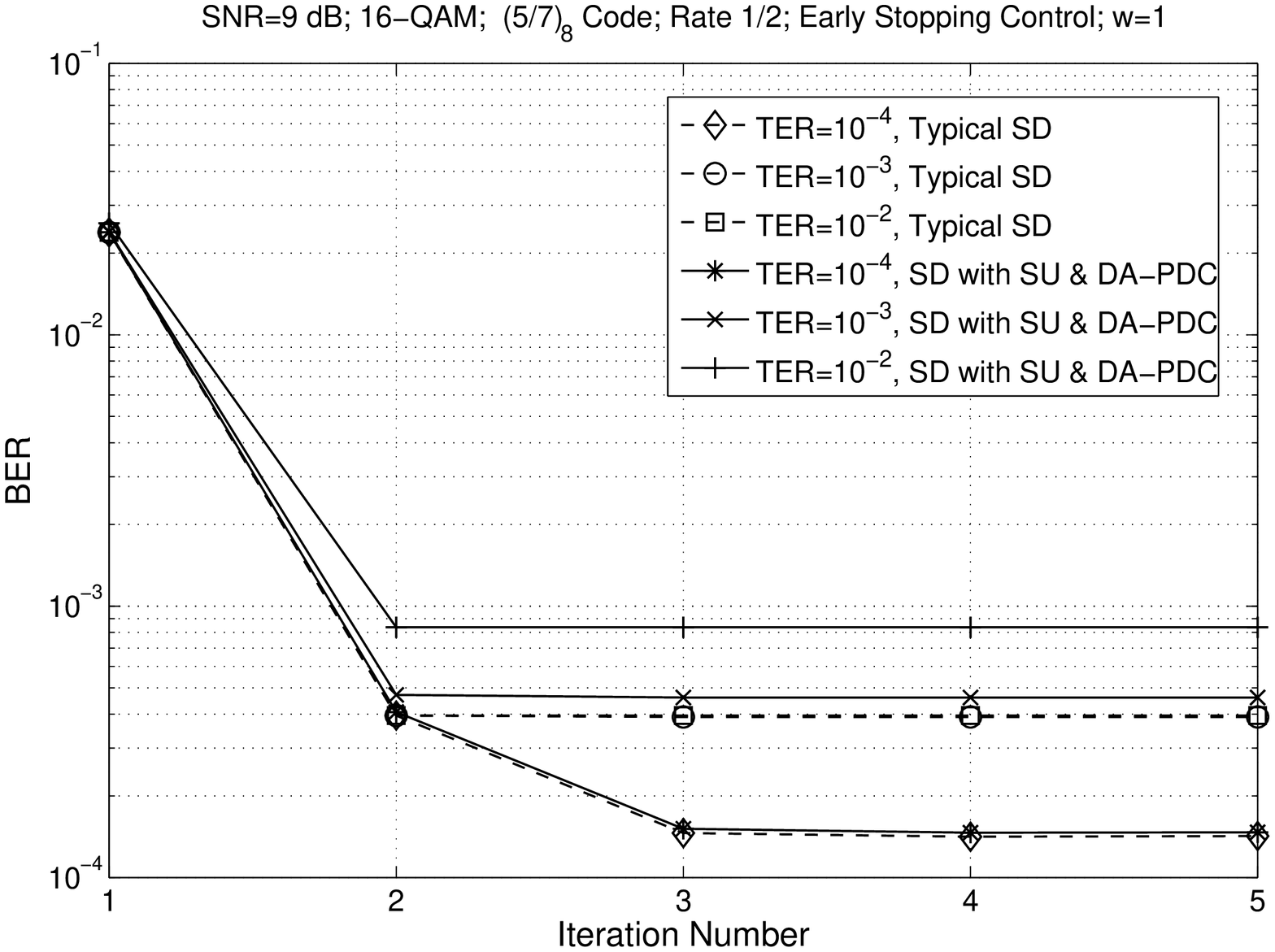}\\
  \caption{BER performance for a system with selective channel decoding of $w=1$,  SU \& DA-PDC soft demapping and several TER values at 9 dB.}
\end{figure}

\newpage
\begin{figure}
  \centering
\includegraphics[width=0.7\linewidth]{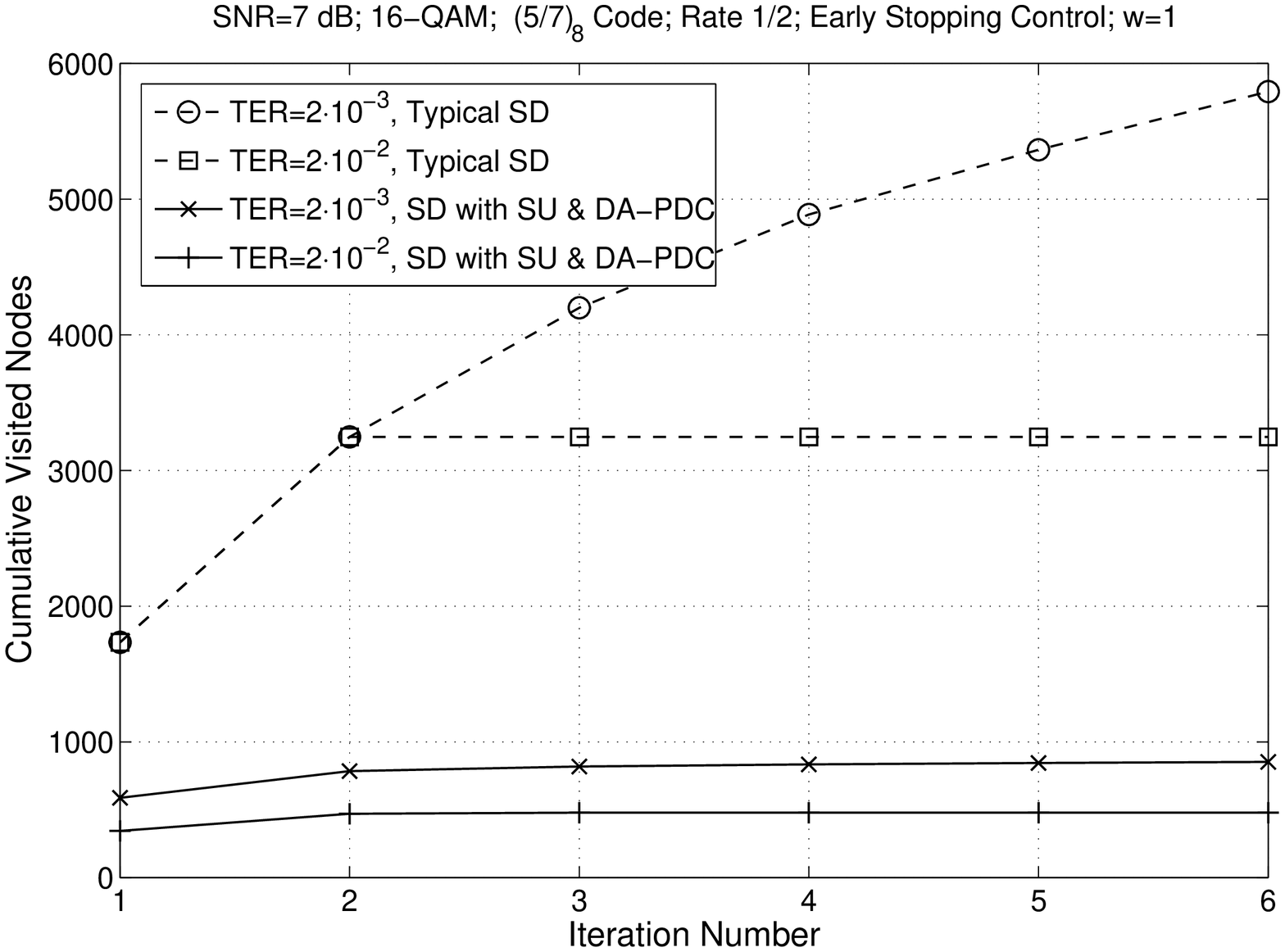}\\
  \caption{Soft demapping complexity for a system with selective channel decoding of $w=1$,  SU \& DA-PDC soft demapping and several TER values at 7 dB.}
\end{figure}

\newpage
\begin{figure}
  \centering
\includegraphics[width=0.7\linewidth]{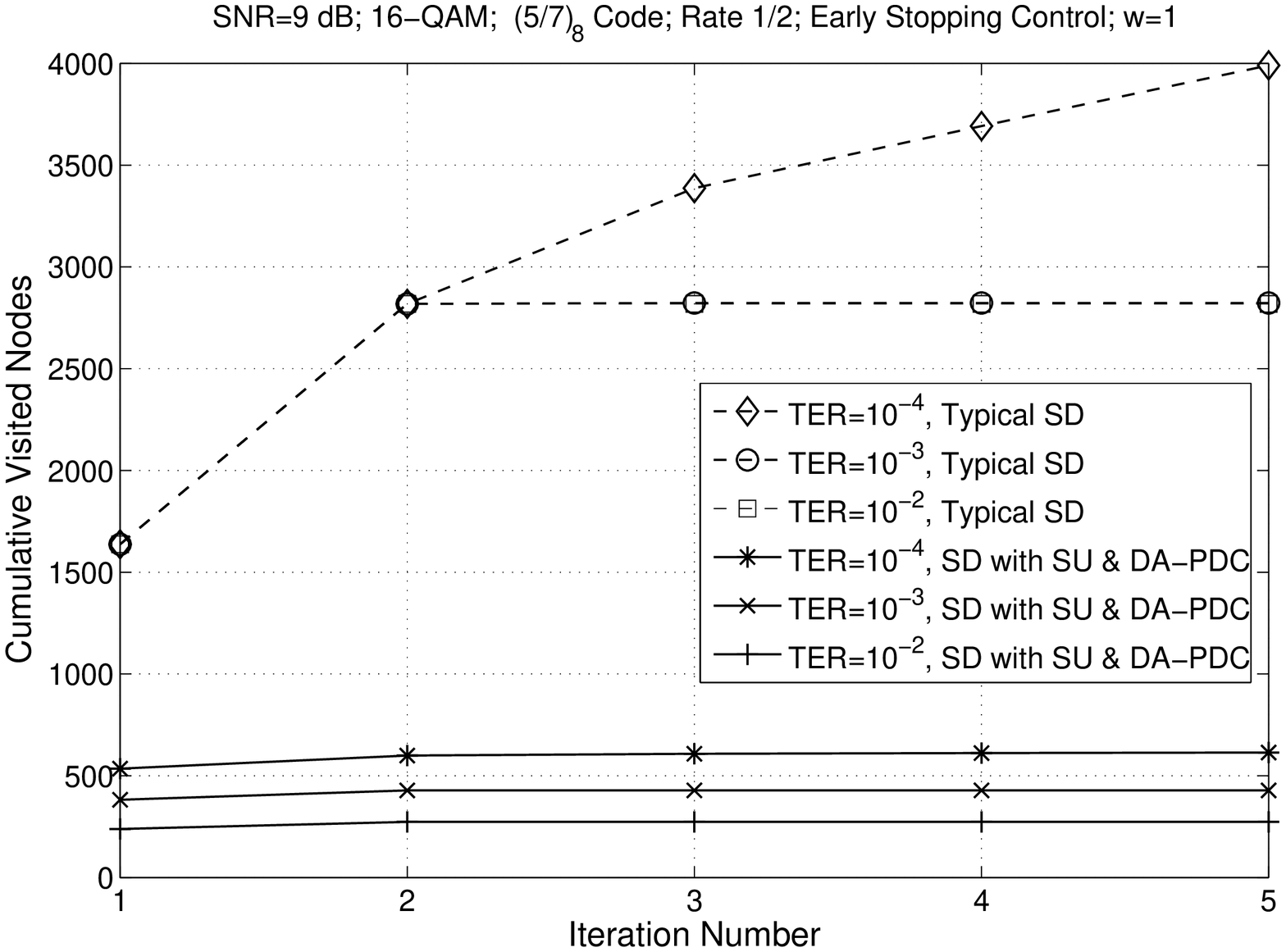}\\
  \caption{Soft demapping complexity for a system with selective channel decoding of $w=1$,  SU \& DA-PDC soft demapping and several TER values at 9 dB.}
\end{figure}

\newpage
\begin{figure}
  \centering
\includegraphics[width=0.7\linewidth]{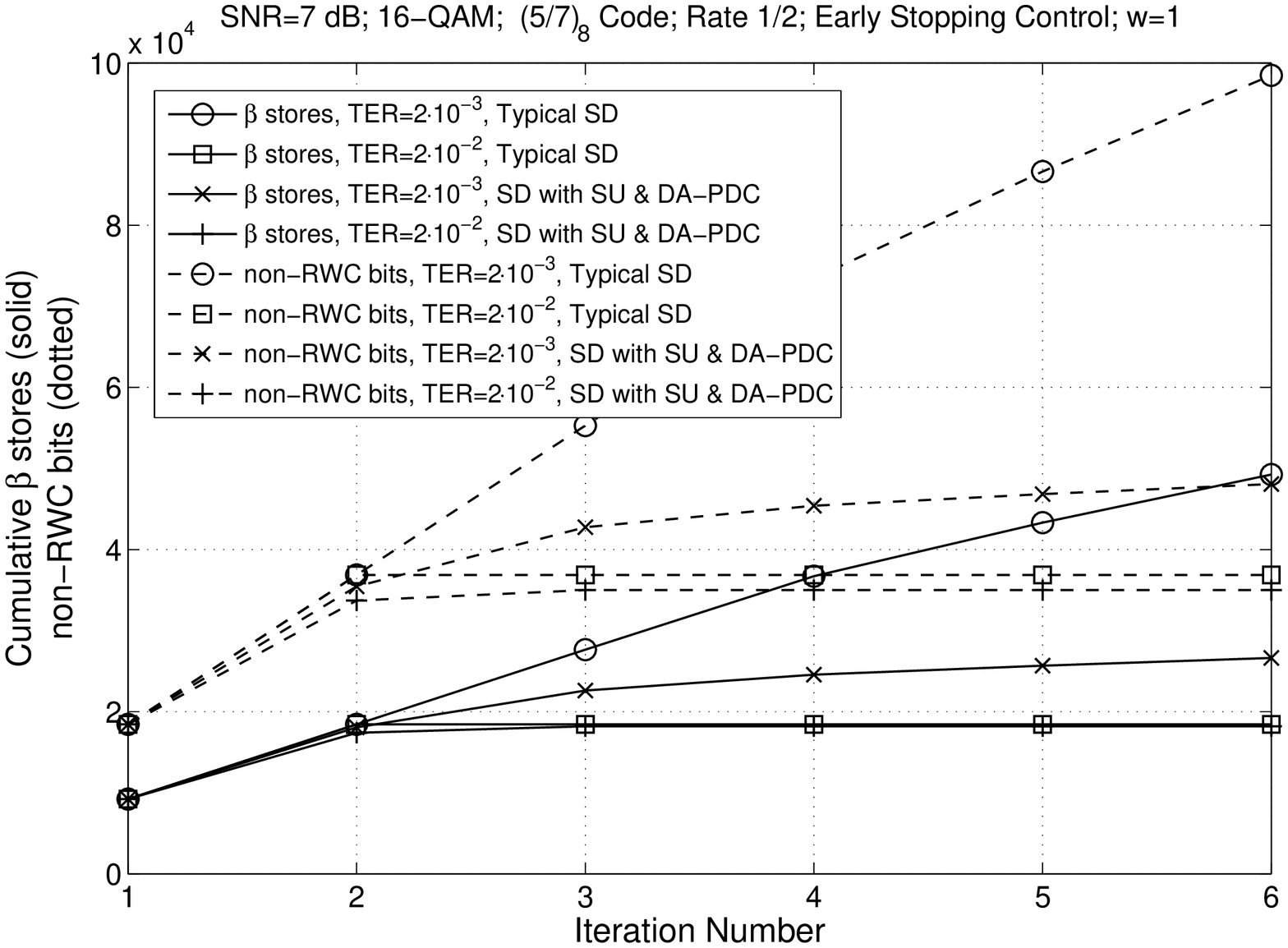}\\
  \caption{Memory store requirement for a system with selective channel decoding of $w=1$,  SU \& DA-PDC soft demapping and several TER values at 7 dB.}
\end{figure}

\begin{figure}
  \centering
\includegraphics[width=0.7\linewidth]{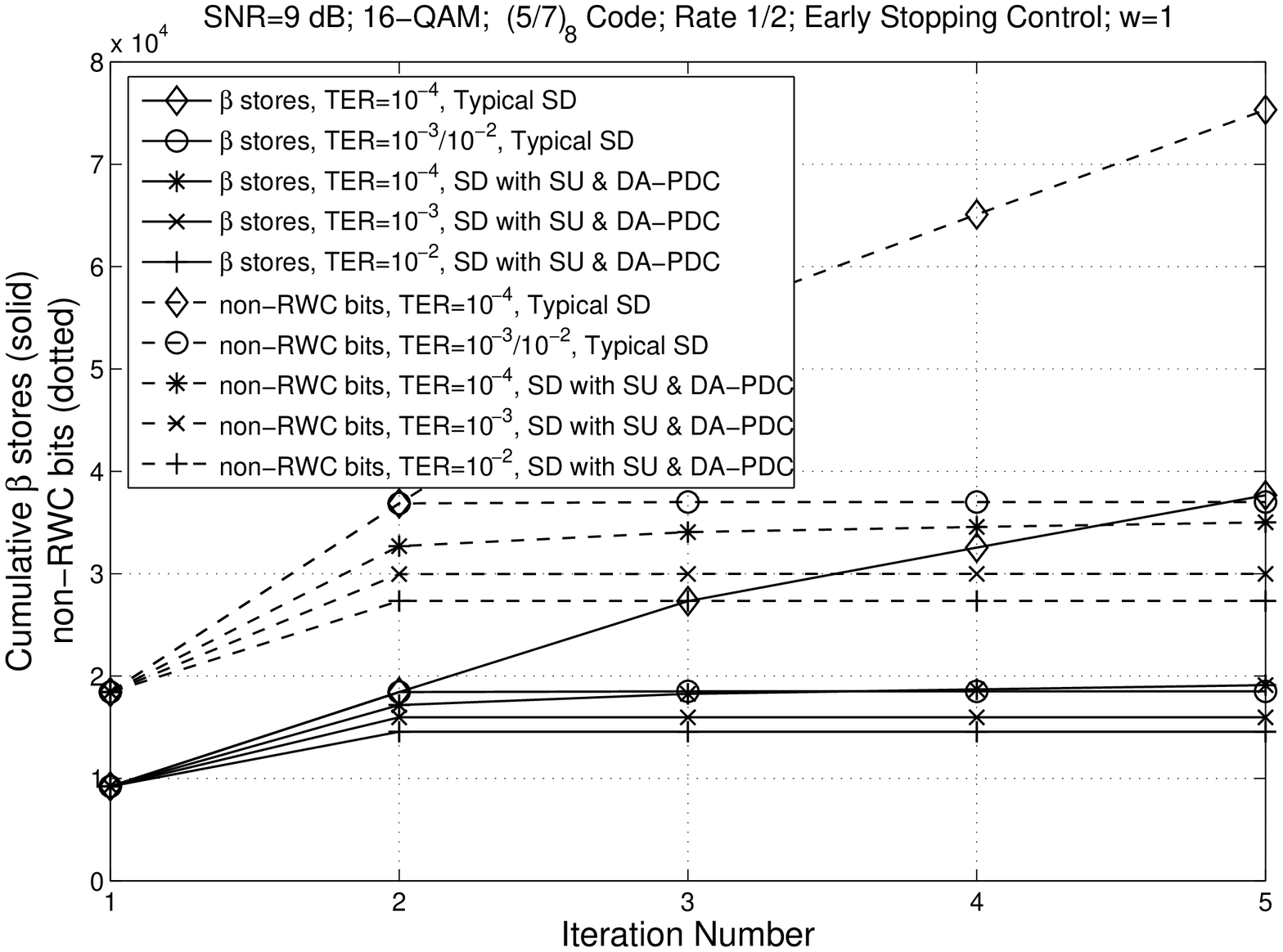}\\
  \caption{Memory store requirement for a system with selective channel decoding of $w=1$,  SU \& DA-PDC soft demapping and several TER values at 9 dB.}
\end{figure}
%\begin{biography}[{\includegraphics[width=1in,height=1.25in,clip,keepaspectratio]{mshell}}]{Michael Shell}

%\end{IEEEbiography}


%\newpage



%\vfill


\end{document}